\documentclass{aa}  
\usepackage{graphicx,tabularx}
\usepackage{txfonts,natbib} 

\begin{document}
   \title{The enigma of GCIRS~3 }

   \subtitle{Constraining the properties of the mid-infrared reference star of the
   central parsec of the Milky Way with optical long-baseline interferometry.\thanks{Based on observations obtained at the European Southern
     Observatory, Paranal, Chile (programs 073.B-0249, 075.B-0113, 077.B-0028).}}

   \author{J.-U.~Pott
          \inst{1}\fnmsep\inst{2}
          \and
          A.~Eckart\inst{1}
	  \and A.~Glindemann\inst{2}
	  \and R.~Sch\"odel\inst{1}
	  \and T.~Viehmann\inst{1}
	  \and M.~Robberto\inst{3}
          }

   \offprints{J.-U. Pott, now at W.M. Keck Observatory: jpott@keck.hawaii.edu}

   \institute{I. Physikalisches Institut, University of Cologne, Z\"ulpicher Str. 77, D-50937 K\"oln, Germany\\
              \email{pott@ph1.uni-koeln.de}
         \and
             European Southern Observatory (ESO), 
              Karl-Schwarzschildstr. 2, D-85748 Garching bei
              M\"unchen, Germany
	 \and
	     Space Telescope Science Institute, 3700 San Martin Drive,
              Baltimore, MD 21218, USA
             }

   \date{Received <date> / Accepted <date>}

 
  \abstract
  {GCIRS~3 is the most prominent MIR-source in the central
  parsec {of the Galaxy}. NIR spectroscopy has failed to solve the enigma of its nature. The properties and peculiarities of extreme individual
  objects in the central stellar cluster contribute to our knowledge
  of star and dust formation close to a supermassive black hole.}
  {We initiated an unprecedented interferometric experiment to understand the nature of GCIRS~3, where we investigate its properties as a spectroscopic and interferometric reference star at 10~$\mu$m.}
  {VLT/VISIR imaging separates a compact source from diffuse,
  surrounding emission. The
  VLTI/MIDI instrument was used to measure spectroscopically resolved
  visibility moduli at an angular resolution of $\sim$ 10~mas of that
  compact 10~$\mu$m source, still unresolved by a single
  VLT. Recent NIR/MIR photometry data were added to enable simple SED- and full radiative transfer-modeling of the data.}
  {The luminosity
  and size estimates show that IRS~3 is   probably a cool carbon star enshrouded by a complex dust distribution. 
Blackbody  temperatures
  were derived. 
The coinciding interpretation of single telescope and interferometric data confirm dust emission from several different spatial scales.
The interferometric
  data resolve the inner rim of dust formation. 
Despite observed deep
  silicate absorption towards GCIRS~3, we favor a carbon-rich
  circumstellar dust shell. The silicate absorption most probably
  takes place in the outer diffuse dust, which is mostly ignored by MIDI
  measurements, but very observable in complementary VLT/VISIR data. This indicates physically and chemically
  distinct conditions of the local dust, changing with the distance to GCIRS~3.}
   {We have demonstrated that optical long baseline interferometry at
  infrared wavelengths is an indispensable tool for investigating sources
  at the Galactic center. Our findings suggest further studies of the
  composition of interstellar dust and the shape of the 10~$\mu$m
  silicate feature in this extraordinary region. }


   \maketitle
%

\section{Introduction}
\begin{figure}
\centering
\includegraphics[width=\columnwidth]{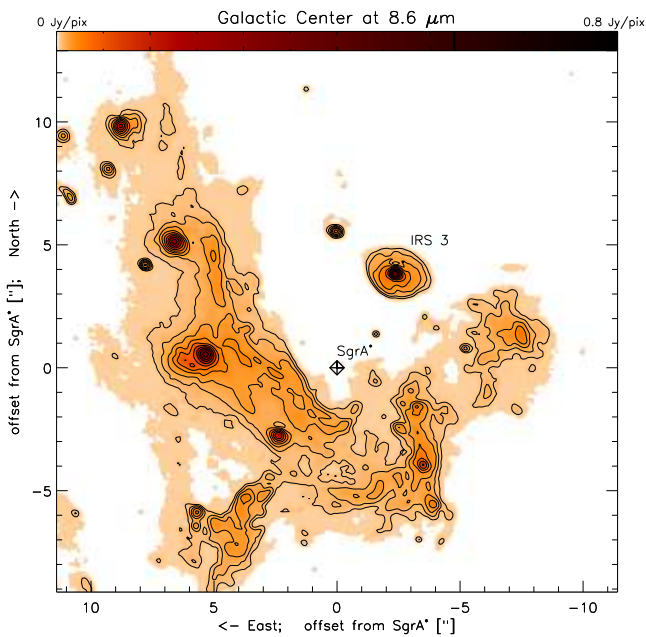}
\caption{Flux-calibrated Lucy-Richardson deconvolved image at
  8.59~$\mu$m after restoration with a 250~mas Gaussian
  beam. The pixel scale is 75~mas per pixel. The logarithmic contours levels are 1.6$^{n}\cdot$7~mJy. The flux scale is
  given on top. IRS~3 and \object{Sgr A*} are highlighted. Details of
  the data reduction are given in \citet{2007A&A...462L...1S}.}
\label{fig:62}
\end{figure}

At a distance of $\sim$7.6~kpc
\citep{2005ApJ...628..246E}, the center of the Milky Way is by far the closest center of a large
spiral galaxy. 
Its astrophysical properties can be studied on a unique angular scale of
$\sim$ 40 mpc/arcsec, which is two orders of magnitude smaller than the
angular scale at the nucleus of \object{M31}, the next comparable
galaxy \citep{2005MNRAS.356..979M}.
Star formation and the kinematics of
the central stellar cluster can be studied in the region of direct
influence of the supermassive black hole (SMBH) at the
dynamic center of the Milky Way
\citep{1996Natur.383..415E,2000Natur.407..349G,2002Natur.419..694S}.

The active history of star formation, despite the tidal forces of the
SMBH, is manifested in the existence of numerous massive, young
stars in the central cluster
\citep{1995ApJ...447L..95K, 2000MNRAS.317..348G,2004ApJ...602..760E, 2004A&A...425..529M}.
The 7 most luminous ($L\,>\,10^{5.75}\,$L$_\odot$), moderately
hot blue supergiants ($T\,<\,10^{4.5}\,$K) provide about 50\% of the
flux ionizing the region
\citep{1995ApJ...441..603B,1995ApJ...447L..95K,1997A&A...325..700N}.
\citet{2003A&A...398..229S} have shown that a few tens of carbon-rich
supergiants can produce about 50\% of the mass loss of a large stellar
sample with solar neighborhood characteristics.
Thus, similar to the ionizing flux, the mass loss and dust-formation properties of a stellar cluster can be 
dominated by a few prominent stars. 
These facts underline the importance of
studying individual extreme objects like \object{GCIRS~3}~\footnote{in the following IRS~3} to understand
properties of the entire surrounding stellar environment.

The recent advent of mid infrared\footnote{3-20 $\mu$m covering the
  atmospheric $L$, $M$, $N$, and $Q-$windows} (MIR) instruments on 8~m class
telescopes enables us to study in detail the thermal dust at the Galactic
center (GC) at unprecedented angular resolution. 
The investigation of the
  circum- and interstellar dust distribution at the GC uncovers
  stellar mass loss, zones of wind interaction, formation history, evolution, and kinematics.

Photometric and spectral properties of dusty stars at the
  GC have recently been published by \citet{2004A&A...425..529M} and \citet{2005A&A...433..117V,2006ApJ...642..861V}. 
Despite an average optical extinction of $A_{\rm V}\,\sim\,$25
  \citep{2003ApJ...594..294S,2005A&A...433..117V}, near-infrared
  spectroscopy and imaging reveal the nature
  of the underlying dust-embedded stars in most cases, since the dust extinction
  is wavelength-dependent and decreases  from the optical towards longer wavelengths.

This article focuses on the most prominent of the MIR bright dusty sources,
  \object{IRS~3}, the
  embedded stellar source, which still
  eludes any spectral classification
  \citep{2006ApJ...643.1011P,2006ApJ...641..891T}. 
  It is located within the central 20''. 
A recent state-of-the-art narrow-band MIR image at 8.6~$\mu$m with an angular resolution of
  only 250~mas is shown in Fig.~\ref{fig:62}. 
The extended and diffuse dust
  emission surrounding IRS~3 is visible at unprecedented angular resolution in this image.

The NIR extinction studies
  reveal a spatial variation of
  only $\sim$10\% ($A_{\rm K}\,\approx\,3$) of the interstellar
  extinction over this region
  \citep{2003ApJ...594..294S,2004A&A...425..529M,2007A&A...469..125S}. 
 
In contrast, narrow-band, $N$-band photometry and spectroscopic
  observations are interpreted to indicate a significant amount of
  {\em additional} 10~$\mu$m silicate absorption along the line of sight towards
  IRS~3 with respect to other parts of the central 20''
  \citep{1978ApJ...219..121B,1985MNRAS.215..425R,2006ApJ...642..861V}.
  The published intrinsic optical depths of about $\tau_{9.8}({\rm
  IRS~3}) \sim 1$, in addition to the average $\tau_{9.8}({\rm
  GC})\,\approx \,3.5$
  \citep[e.g.][]{1985MNRAS.215..425R}, still underestimate the true
  value due to source confusion problems. While our new high-resolution VISIR data clearly
  indicate that more than 50\% of the $N$-flux is diffuse, extended
  emission at 0.3'' resolution, one byproduct of the MIDI observations
  is the calibrated, low-resolution spectra of the compact emission, which show $\tau_{9.8}$(IRS~3)
  to be much larger than the value given above.

\citet{1978ApJ...220..556R} found the spectro-photometric MIR
  properties of IRS~3 to resemble either young stars or OH/IR stars. 
The latter interpretation is opposed by the lack of OH-maser emission.
With an MIR color temperature of $\sim$~400~K, the central emission of IRS~3 was found to be
  (together with the nearby GCIRS~7) the hottest and most compact of
  the sources dominating the thermal dust irradiation from the GC \citep{1978ApJ...220..556R,1985ApJ...299.1007G,1990MNRAS.246....1S}.
Extended dust emission around IRS~3 interacting with external
  stellar winds has been found in recent $L-$ and $M-$band observations \citep{2005A&A...433..117V}.
While hot star hypotheses are given by some authors
  \citep{1995ApJ...447L..95K,2003ANS...324..597T}, the lack of ionizing gas leads \citet{1985MNRAS.215..425R} to the
  assumption of IRS~3 being a cool dust-enshrouded star.

Within the past two years we collected a unique dataset of optical
long-baseline interferometric (OLBI) data of IRS~3 at 10~$\mu$m. 
These constitute the first successful OLBI observation of an object
within the central parsec of our galaxy, opening the window to NIR/MIR
GC observations at highest angular
resolution \citep{2005Msngr.119...43P}.
  In this article we show that the OLBI data strongly support the
former hypothesis of a cool dust-enshrouded star. Furthermore our
results shed light on the amount and spectral shape of the
interstellar 10~$\mu$m extinction towards IRS~3 at unprecedented
angular resolution. 
Since MIDI is a relatively new instrument, the
achievable precision of visibility measurements of such distant, challenging
targets is not common knowledge yet. 
In Sect.~\ref{sec:2} we therefore describe the
observations, the extensive data-reduction and calibration efforts,
and the evaluation of different data reduction techniques in some detail to
ease and stimulate similar experiments and the comparison of their outcomes.
Then the immediate observational results are given
(Sect.~\ref{sec:30}), followed by a detailed discussion of the results in the astrophysical context (Sect.~\ref{sec:4}) and a summary of our conclusions in Sect.\ref{sec:6}. 


\section{Observations and data reduction \label{sec:2}}

\begin{table}
\caption{Observing log of IRS~3. The wavelength
  dependent $TF$ has been estimated for each night on
  the basis of regularly conducted calibrator measurements, typically
  about once per hour (Sect.~\ref{sec:2}). The number of calibrators used per night is
  given, and the applied parameters are given in Table~\ref{tab:2}. 
}
\label{tab:1}      
\centering                          
\begin{tabular}{c c c c c}        
\hline\hline                 
 Baseline & PB & PA & Airmass & Accuracy\\    
  & [m]& [deg E of N] & [1] & $[$\%$]$\\
\hline
\multicolumn{4}{l}{\#1 night: 2004-07-07 \quad - \quad 5 calibrators} \\
U2-U3 & 45.7 & 46 & 1.03 & 12\\
U2-U3 & 42.3 & 52 & 1.16 & 12\\
U2-U3 & 37.0 & 55 & 1.45 & 12\\
U2-U3 & 26.4 & 53 & 2.63 & 14\\
\hline
\multicolumn{4}{l}{\#2 night: 2004-07-08\quad - \quad 8 calibrators} \\
U2-U3 & 44.9 & 11 & 1.38 & 19\\
U2-U3 & 43.5 & 50 & 1.11 & 14 \\
\hline
\multicolumn{4}{l}{\#3 night: 2005-05-25\quad - \quad 7 calibrators} \\
U3-U4 & 45.8 & 89 & 1.36 & 14\\
\hline                                   
\multicolumn{4}{l}{\#4 night: 2005-06-23\quad - \quad 4 calibrators} \\
U3-U4 &  45 & 89 & 1.39 & 23\\
U3-U4 &  46.8 & 90 & 1.33 & 16\\
\hline                                   
\multicolumn{4}{l}{\#5 night: 2005-06-27\quad - \quad 4 calibrators} \\
U3-U4 & 54& 139 & 1.29 & 11\\
\hline                                   
\multicolumn{4}{l}{\#6 night: 2005-07-20\quad - \quad 7 calibrators} \\
U3-U4 & 58.6 & 101 & 1.07 & 22\\
U3-U4 & 59.6 & 124 & 1.08 & 22\\
\hline                                   
\multicolumn{4}{l}{\#7 night: 2005-08-23\quad - \quad 2 calibrators} \\
U3-U4 &  62.5& 112 & 1.00 & 20 \\
U3-U4 &  54.9& 136 & 1.25 & 21 \\
\hline                                   
\end{tabular}
\end{table}

\begin{table}
\begin{minipage}[t]{\columnwidth}
\caption{Parameters of the calibrators used for estimating the
  transfer function. }
\label{tab:2}
\centering
\renewcommand{\footnoterule}{}  
\begin{tabular}{ccccc}
\hline\hline
Name & Spectral & Diameter & Flux dens. & Night \footnote{The night in which the  calibrator was used}\\
 & type & [mas] & [Jy $@$12$\mu$m] & \\
\hline
HD107446 & K3.5III &4.42$\pm$0.03& 22.0 & 4 \\
HD109379 & G5II & 3.25$\pm$0.02 & 15.1 & 1\\  
HD123139 & K0IIIb & 5.34$\pm$0.03 &36.4& 3\\
HD134505 &G8III& 2.50 $\pm$0.01& 8.4 & 4 \\
HD142804 &M1III & 2.79$\pm$0.04& 8.0& 7\\
HD152820 &K5III & 2.62$\pm$0.01&7.5& 3,4\\
HD160668 &K5III & 2.28$\pm$0.01 & 5.7 & 3,5\\
HD165135 &K0III & 3.48$\pm$0.02 & 15.5 & 1-7\\
HD169767 &G9III & 2.16$\pm$0.01 & 5.9 & 3 \\
HD169916 &K1IIIb & 4.00$\pm$0.03 &20.1 & 4 \\
HD173484 &M4III & 3.43$\pm$0.04 & 11.9& 5,6\\
HD177716 &K1IIIb & 3.76$\pm$0.03& 17.1& 2\\
HD178345 &K0II & 2.49$\pm$0.01 & 7.6& 5-7 \\
HD188512 &G8IV & 2.07$\pm$0.01 & 5.9& 1-3\\
HD192947 &G6/G8III & 2.33$\pm$0.02 &7.5& 1\\
HD213009 &G7III & 2.08$\pm$0.02 & 5.9 & 5 \\
HD218594 &K1III & 3.18$\pm$0.02 & 12.4& 3 \\
HD220704 &K4III & 3.45$\pm$0.02 & 13.2 & 7 \\
\hline
\end{tabular}
\end{minipage}
\end{table}

   \begin{figure}
   \centering
   \includegraphics[width=\columnwidth]{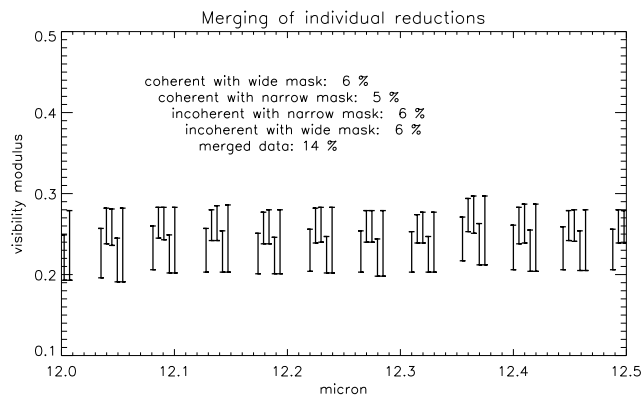}
      \caption{Part of the science data taken on 2005-05-25. We overplot the different
              reduction methods (coherent and
              incoherent fringe averaging and static and dynamic
              detector apertures) to demonstrate the different results.
The labelling indicates the artificial wavelength shifts successively applied to increase the readability of the plot.
The relative errors of each method is given. 
The rightmost error bars present
              the merged, maximum error bars. We based our analysis on these
              merged data. 
              }
         \label{fig:1}
   \end{figure}


In 2004 we started an observing campaign to study the brightest MIR-excess sources in the central parsec with MIDI at the highest angular resolution available today. 
The MID-Infrared interferometric instrument (MIDI) combines the light of two 8.2m unit telescopes of the ESO Paranal Observatory in Chile \citep{2003Msngr.112...13L}. 
We used the standard 0.5''$\times$2'' slit and dispersed the light over the entire $N-$band (8-13~$\mu$m) with the prism providing a spectral resolution of $R\sim 30$.
The first MIR fringes of IRS~3 were recorded successfully
on the night of 7~July~2004.
The whole dataset comprises 14 independent visibility measurements of
IRS~3 with interlaced calibration measurements. 
The projected baseline length (PB) and position angle (PA) of each measurement are given in Table~\ref{tab:1}
  for the beginning of the fringe measurement. The last column gives
  the median relative uncertainty obtained for the calibrated
  visibility outside the low-flux region of the silicate absorption.

\subsection{\label{sec:21} Interferometric calibrator stars}

In Table~\ref{tab:2} we list the used calibrators with their main features.
The angular diameter ascertainments result from fitting {\sc Atlas9} and
{\sc Marcs} model atmosphere spectral energy distributions (SEDs) to optical and NIR photometry.
The chosen models \citep{1992IAUS..149..225K,1994KurCD..19.....K,1992A&A...256..551P} adopt solar metallicity. 
Details of the model fitting are given in \citet[][chapter 5]{2004PhDT.........V}.
The modeled 12$\mu$m flux densities listed in Table~\ref{tab:2} are consistent within $\lesssim$5\% with the work of \citet{1999AJ....117.1864C}, who present a list of absolutely calibrated infrared spectra. 
Furthermore, none of those calibrators shows an MIR-excess, defined as having a measured 12 $\mu$m flux density more than 3~$\sigma$ above the synthetic spectra fitted to the optical-NIR data.
Such an MIR-excess would indicate the existence of (extended) dust shells. 
Dust shells, which can be expected to exist around K-M giants, radiate
the stellar flux at longer wavelengths and decrease the visibility
amplitudes of the entire object; i.e., they deteriorate the calibrator properties. 
The absence of an MIR-excess also makes all visibility calibrators usable as potential photometric calibrators, down to the 5\% uncertainty, which is often not reached due to atmospheric and instrumental variability.
The photometric variability and the fitting uncertainty of the angular diameter of all the calibration stars in Table~\ref{tab:2} affect the accuracy of the derived transfer functions by less then 1\% at all VLTI baselines ($\le$ 200~m).

\subsection{\label{sec:22}Calibration and absolute accuracy}
\label{sec:11}
\begin{figure}
\includegraphics[width=\columnwidth]{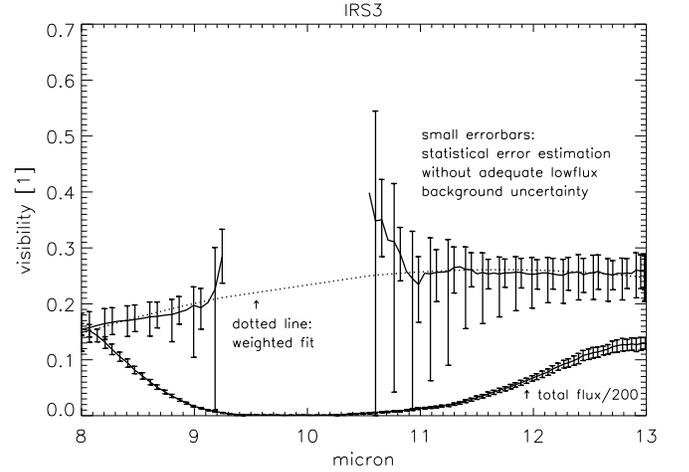}
\caption{The influence of faulty background estimation on the final
  accuracy for data from 2005-05-25.   The
   standard reduction is plotted (solid line) with respective uncertainties (small errorbars to the right of larger ones).
It suggests a significant
  deviation from a smooth, flat visibility spectrum (dotted line: weighted quadratic fit over the silicate absorption
  feature). In particular the visibility increase (solid line) at $\lambda \le$~11~$\mu$m towards the absorption center cannot be confirmed if we
  consider the impact of background subtraction errors (large error bars). 
To demonstrate the flux dependence of the importance of
  background accuracy, we overplotted the total flux spectrum (in Jy).}
\label{fig:3}
\end{figure}

\begin{figure}
\centering
\includegraphics[width=0.95\columnwidth]{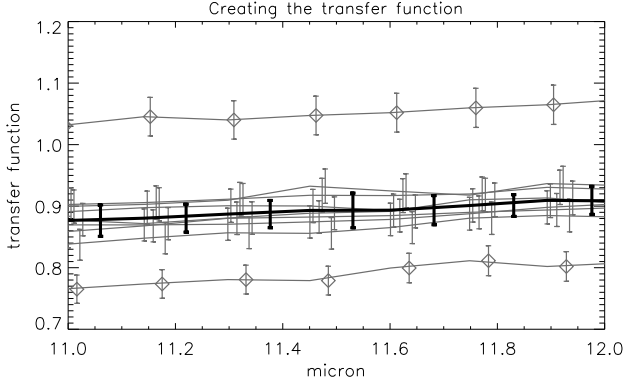}
\caption{Transfer functions in gray, derived
  from the individual calibrator measurements of 2005-05-25. The error
  bars around 5\% indicate the statistical error of the single
  data frames of one measurement, which is slightly smaller than the absolute variation in the $TF$ over the night. The mean $TF$, which we used for this night, is overplotted in
  bold solid black style with the standard deviation. 
 The two remote $TF$ (gray), highlighted with a diamond symbol, were rejected from calculating the mean
  because of abnormal behavior relative to the average of the good
  $TF$. }
\label{fig:11}
\end{figure}

\begin{figure}
\centering
\includegraphics[width=0.95\columnwidth]{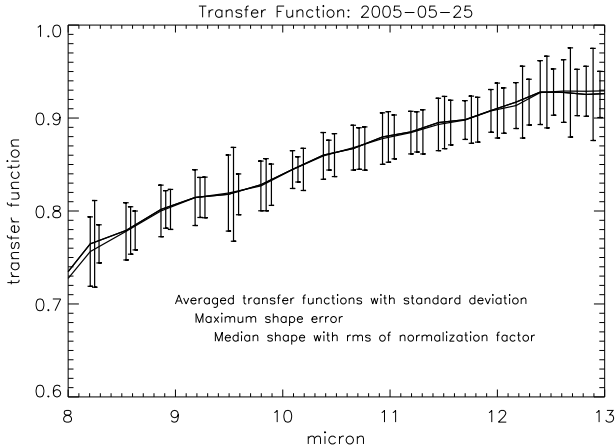}
\caption{Different analyses of the $TF$, derived from
  calibrator measurements on 2005-05-25. To increase
 the clarity of the plot, the wavelengths of two overplots have been shifted, as
  indicated by the respectively shifted labeling. Left error bars: average (mean \& standard deviation) of the
  individual $TF$. The two other data sets are averaged in normalized
  space (division of the individual $TF$ by their
  integral). Central: maximum differential
  fluctuation between adjacent spectral channels, presenting a
  conservative maximum of the spectral shape error of $TF$. Right: standard deviation of the
  normalization factor, representing the order of magnitude of
  the variability of transmission between the $F_{\rm C}$ and $F_{\rm
  T}$ measurement. }
\label{fig:10}
\end{figure}

The immediate measurand, the raw visibility $V_{\rm raw}$, is the ratio of measured correlated ($F_{\rm C}$) and total flux ($F_{\rm T}$). 
The final quantity of interest is the calibrated visibility ($V_{\rm cal}$), which is computed by dividing the raw visibility ($V_{\rm raw}$) by the interferometric transfer function ($TF$) of the observation.
Propagation of errors relates the uncertainties $\Delta$   to each other:
\begin{eqnarray}
\label{equ:1}
V_{\rm calib}= \frac{V_{\rm raw}}{TF} \hspace{1cm} \rightarrow  \left(\frac{\Delta V_{\rm calib}}{V_{\rm calib}}\right)^2&=&\left(\frac{\Delta V_{\rm raw}}{V_{\rm raw}}\right)^2+ \left(\frac{\Delta TF}{TF}\right)^2
\\
\label{equ:2}
\left.TF= \frac{V_{\rm raw}}{V_{\rm ideal}}\,\right|_{\rm calibrator} \rightarrow \hspace{0.2cm} \, \left(\frac{\Delta TF}{TF}\right)^2 &=& \left.\left(\frac{\Delta V_{\rm raw}}{V_{\rm raw}}\right)^2\,\right|_{\rm calibrator}.
\end{eqnarray}
{The $TF$ is derived from the $V_{\rm raw}$ measurement of a calibrator
star and its intrinsic visibility (Eq.~\ref{equ:2}). The uncertainty
of the diameter of the calibrator is usually too small to affect $\Delta TF$ (Table~\ref{tab:2}).}
Thus $\Delta V_{\rm calib}$ suffers twice from the accuracy of the
estimation of $V_{\rm raw}$ (Eq.~\ref{equ:1}). 
This   central accuracy can be   estimated by means of the
  following considerations and investigations of calibrator data sets.

All constant defects of $V_{\rm raw}$ are corrected for by
  multiplication with $TF$ and thus have only a minor influence on the final accuracy; e.g., the time delay between the measurement of $F_{\rm C}$ and
$F_{\rm T}$ is only used a few minutes in the high-sensitivity mode of MIDI.
This reduces the variation of atmospheric transmission in the thermal
infrared, due to airmass difference and temporal fluctuations, to such a
  level that a small and time-constant impact on the quotient $F_{\rm
  C}$/$F_{\rm T}$ can be assumed.
  We prove this by the following test, visualized in
  Fig.~\ref{fig:10}. If the atmospheric and instrumental flux
  transmission change randomly at a significant level between the $F_{\rm
  C}$ and $F_{\rm T}$ measurements, the orders of magnitude of $V_{\rm raw}$ and the resulting
  $TF$ are affected and the absolute accuracy of the calibrated
  visibility is worse than the relative spectral accuracy.
To derive this spectral accuracy we linearly normalize each individual $TF$ of
  a night by dividing through the integral over a fixed part of the
  $N$-band and estimate the maximum scatter in each spectral
  channel. If this scatter is significantly smaller than the scatter
  of the normalization factors, a strong transmission variability has
  been observed (assuming constant intrinsic
  visibility). Figure~\ref{fig:10} shows that
  this is not the case. Note in the same figure the increase in the shape error around the
  atmospheric ozone absorption band at 9.7~$\mu$m and at the borders
  of the $N$-band, where the applicability of simple linear normalization decreases.

  Faulty and unequal background suppression during reduction of both
data sets also does not vary on a significant level for the bright
  calibrator stars. The influence of background noise for
  faint science targets is discussed at the end of the section.

As a matter of fact, \citet{2004A&A...423..537L} state that the accuracy of
$V_{\rm raw}$ is dominated by the accuracy of the overlap between the
interfering beams, which critically relies on the wavefront
corrections during observations.
This is difficult to quantify and   thus cannot be easily incorporated into the $TF$. Usually   the quality of the
  beam overlap does not vary {\em statistically} during one observation.
This means that there is no {\em mean} overlap accuracy to correctly describe a
  single observation and that the $TF$, if derived from one calibrator
  measurement, may not be applicable to a subsequent science observation.
We confirm this on the basis of our dataset. 
A single observation of both
$F_{\rm C}$ and $F_{\rm T}$ consists of
several frames that are averaged during data reduction. 
We estimate
the standard deviation of the subsequent scans and analyze the
background in chopped photometry data to derive a statistical
uncertainty for each measurement of $V_{\rm raw}$.
  The resulting total statistical uncertainty $\sigma (V_{\rm
    raw})$ is similar for the individual calibrator observations, but it
cannot fully explain the larger scatter of individually
estimated $TF$ over the respective observing night. 
Figure~\ref{fig:11} demonstrates that the standard deviation over several calibrators is only as small as expected from the estimated uncertainty of the individual measurements ($\sim$~5\%), if outliers are rejected from the average.
That means that the absolute accuracy of the visibility measurement
can change significantly with each new pointing. 

Furthermore, the overnight scatter of $TF$ is significantly reduced in the 2005
data due to the more stable VLTI feeding by the higher-order AO-system
{\sc Macao} instead of the earlier tip-tilt-only correction of the
{\sc Strap} units \citep{2003SPIE.4839..174A}. This also confirms the
dominating influence of varying beam overlap and flux concentration
between different pointings on the visibility accuracy.

The above given tests and considerations underline the origin of the
accuracy of the $V_{\rm raw}$ measurement and justify the accuracy of a single measurement only being given by the
$TF$ statistics of {\em several} calibrator measurements.
We calculated the mean and standard deviation of the different $TF$ to quantify $TF$ and $\Delta TF$.
  Note that the presented estimation of errors even holds for an extended
calibrator with visibilities significantly lower than one, provided
that the diameter of the calibrator is known at sufficient precision. Although $V_{\rm
  raw,\,calibrator}$ will systematically change over the night with changing projected baseline length, this effect is annihilated by calculating $TF$ (Eq.~\ref{equ:2}).

Since the number of calibrators per night is typically around 8, one  $TF$ strongly deviating from the others can influence the mean and standard deviation significantly. 
To avoid this situation we did not use the median, but rejected the anomalous $TF$ manually (on average one per night, Fig.~\ref{fig:11}).
This has the advantage that the mean and standard deviation of the
resulting sample better represents the spectral shape of the $TF$
reducing the probability of artificial spectral features in the calibrated visibility spectrum.

To complete Eq.~\ref{equ:1}, we have to know the relative uncertainty of $V_{\rm raw}$ of the measured target. 
Qualitatively the origin is similar, as discussed above for the calibrators,
but the flux of the science targets can be significantly smaller than the calibrator fluxes. 
Since our calibrators cover a range of 5-35~Jy, we searched for any flux dependence of $\Delta V_{\rm raw}$ on the basis of our data.
We checked for every night whether on average the scatter of the $TF$ of
the fainter calibrators is greater than the scatter of the $TF$ of brighter ones, but did not find any such flux dependence.
Thus we assume a flux-independent relative uncertainty of $V_{\rm raw}$ resulting in the uncertainty of the calibrated visibility:
\begin{eqnarray}
\label{eqn:1}
\frac{\Delta V_{\rm calib}}{V_{\rm calib}}&=& \sqrt{2}\cdot\frac{\Delta TF}{TF}\quad .
\end{eqnarray}
  This estimation might still underestimate the relative accuracy of the
final calibrated visibility of the science target, since IRS~3 was observed with an off-axis
AO-guide star, which decreases the AO performance with respect to the on-axis guiding on the calibrators. We compared the PSF
of the photometric data of IRS~3 with the calibrator measurements and
specified three datasets with calibrator-like PSF. These are
highlighted throughout the analysis and provide the tightest
constraints on our interpretation of the data (cf. Sect.~\ref{sec:24}~\&~\ref{sec:30}).

If the correlated flux drops significantly below  5~Jy and
approaches the sensitivity limits, an increase in the relative
uncertainties will probably occur due to the increased influence of
noise; but outside the deep silicate absorption, the correlated spectrum of IRS~3 is above this limit.
Furthermore, imperfect background subtraction has an increasing
influence on the final accuracy with decreasing source flux. 
To account for this, we estimated the background level by reducing
sky-frames without the source.
Typically the background randomly varies between zero and a value
close to $F_{\rm bg}$, where $F_{\rm bg}$ is the sum of the mean and
standard deviation of all spectral channels of the background frames.
This leads to a maximum background induced error interval of
\begin{eqnarray}
V_{\rm raw, bg} \in \left( \frac{F_{\rm C}-F_{\rm C, bg}}{F_{\rm
    T}},\,\frac{F_{\rm C}}{F_{\rm T}-F_{\rm T, bg}}\right) \quad , 
\end{eqnarray}
which dominates the flanks of the strong silicate absorption
towards IRS~3.
A putative increase in visibility towards the absorption center, which
appears to be present after applying the standard reduction, cannot
be verified after incorporating the background uncertainty (Fig.~\ref{fig:3}). Indeed, not all
  datasets show such an increase after the standard
  reduction. 

\subsection{Data reduction}
\label{sec:23}

We used the data reduction package {\sc Mia+Ews} provided by
the MIDI consortium\footnote{See the ESO web pages for general information  and
  links ({\tt http://www.eso.org/instruments/midi/}). Currently the
  data reduction package is provided on: {\tt
    http://www.strw.leidenuniv.nl/$\sim$nevec/MIDI/}.}.
This package offers two different methods for reducing the data: {\em incoherent} averaging of the fringe power in each spectral bin over several scans, and {\em coherent} averaging of the single dispersed scans. 
In the $N-$band, the latter method can provide the differential phase
information  in addition to the visibility amplitude, if the atmospheric and instrumental delay and dispersion have been removed properly. 
A more detailed description of both methods and their realization in {\sc Mia+Ews} is given by \citet{2004A&A...423..537L} and \citet{2004SPIE.5491..715J}, respectively. 
  Since noise always contributes positively to the power spectrum,
  it is not automatically reduced by averaging incoherently over
  several scans. In contrast, the coherent integration reduces the
statistical noise of the fringe data by averaging, which makes this
method favorable at very low correlated flux (below about 1~Jy)   and
low SNR. 
Correlated fluxes down to 0.1~Jy could be estimated by this method (Jaffe, private communication).
Since our MIDI data is usually well above this limit, we found consistent results of both reduction algorithms throughout the complete dataset (Fig.~\ref{fig:1}). 

The constantly changing baseline projection due to earth rotation
limits the integration times of the $T_C$ measurement.  
  To reduce the noise
  level, which is intrinsically high in the thermal infrared, only the pixels with the highest SNR should be considered by the
  reduction algorithm. This is achieved by detector aperture masks,
  which  can lead to reduced SNR if the chosen aperture is too large,
too small, or misplaced with respect to the incident source photons.
This effect is strongly enhanced by beam distortions and motions that are only partially corrected by the AO system, which
holds especially for our data since we had to lock the AO on an off-axis guide star 35'' away from the GC.
Therefore we reduced the full dataset four times: with both coherent
and incoherent fringe averaging and by applying two different
apertures (or detector masks), a wide standard one with fixed location and width
and a narrower one that is chosen dynamically for the PSF of every
observation to trace the beam maximum and to
have an optimized width (this is the {\em narrow} mask in Fig.~\ref{fig:1}).

After calibrating the data and estimating its accuracy as described in Sect.~\ref{sec:22}, the four spectra $(V_{\rm calib,1..4}\pm \Delta V_{\rm calib,1..4})$ were averaged.
To estimate the accuracy conservatively, we selected a maximum $\Delta
V_{\rm calib}$ including all four uncertainty intervals
(Fig.~\ref{fig:1}).

Note that the apparent correlation between
              systematically higher visibilities and the use of the
              narrower mask in this figure is not a general systematic effect,
              but depends on the data set. Other data show an
              apparent correlation between visibility and averaging 
              method, independent of the chosen mask. Consequently
              we do not prefer one single reduction method over
              another but use the merged data. Furthermore, the
              differences are {\em significant } with respect to the
              intrinsic uncertainties, resulting in clearly increased
              uncertainties of the merged data.
The mean final relative accuracy of each merged scientific dataset is given in Table~\ref{tab:1}.

\subsection{Photometric Calibration}
\label{sec:24}
MIDI observations provide the astronomer  with spectra of the total
flux of the target in addition to the visibility
modulus.
We flux-calibrated the spectra on the basis of the regularly observed
interferometric calibrator stars, typically late type giants
(Sect.~\ref{sec:21}).
We fitted an airmass-dependent system response to the calibrator
measurements of each night to flux-calibrate the science
data \citep[for a more detailed description see][]{2005A&A...437..189V}. 
Spectra that were obviously faulty were not taken into account. 
Such a time-independent system response model makes it possible to use all the good calibrator measurements of the night, which is especially favorable if the broad band spectrum of the science target is not known.
Furthermore, the airmass dependence can minimize the impact of larger
distances between the calibrators and the target. 
In some nights only one star was closer than 10$^\circ$ to the target because of scheduling requirements.
On normal nights, incorporating airmass-dependence reduced the
calibration uncertainties by up to  5\%.
But then the model includes uncertainties due to varying atmospheric
transmission and the instrumental throughput over the night, resulting
in final photometric accuracies of about 5-10\%, which dominate the intrinsic uncertainties of the used calibrator spectra.

The Gaussian detector masks (Sect.\ref{sec:23}) that are typically applied do not affect the visibility calculation but
only the photometry.
If the science target is not completely unresolved by the single
telescope PSF (Fig.~\ref{fig:60}) or
if such a weighting mask is not well-centered on the brightest pixels,
the measured flux is decreased. 
We reduced the photometry separately without applying any mask to take
care of any such bias. 
Furthermore, it turned out that a lot of datasets on the target have at least one
beam of significantly lower quality than the other one. 
Since the calibrator measurements do not show such a strong beam
variation, this effect is assumed to result from the use of an
off-axis AO guide star during the observations of IRS~3, decreasing the accuracy of
the wave-front correction (Sect.\ref{sec:23}).
A manual selection of good datasets facilitates a final
photo-spectrometric accuracy of less than 10\%, which is decreased towards low fluxes due to the remaining background.
The result is shown in the lower spectrum in  Fig.~\ref{fig:33}). 

\section{Results}
\label{sec:30}
In this section we present the direct results of our interferometric
measurements.  First the measured and calibrated data are
  shown. Then in the following sections we discuss how the observed spectra constrain the underlying brightness distribution. This discussion outlines the average properties of IRS~3 in the MIR at MIDI resolution.
\subsection{Visibility moduli}

\begin{table}
\begin{minipage}[t]{\columnwidth}
\caption{Measured mean visibility moduli below and above the
  silicate absorption, centered at 9.8~$\mu$m. }
\label{tab:3}
\centering
\renewcommand{\footnoterule}{}  
\begin{tabular}{ccccc}
\hline \hline
Julian date & PB \footnote{\label{foot:35}PB and PA stand for projected baseline length and position angle and characterize the interferometric resolution at the time of the observation.}& PA \textsuperscript{\ref{foot:35}} & \multicolumn{2}{c}{Visibility} \\
\hline
&  [m]& [deg E of N] & 8-8.7~$\mu$m & 11.8-13~$\mu$m\\
\hline
\multicolumn{5}{l}{\quad \#1 night: 2004-07-07} \\
2453194.7  & 45.7 & 46 & 0.17 $\pm$ 0.05 & 0.27 $\pm$ 0.04\\
2453194.7  & 42.3 & 52 & 0.20 $\pm$ 0.06 & 0.28 $\pm$ 0.04\\
2453194.8  & 37.0 & 55 & 0.24 $\pm$ 0.07 & 0.34 $\pm$ 0.04\\
2453194.9  & 26.4 & 53 & 0.47 $\pm$ 0.19 & 0.53 $\pm$ 0.10\\
\hline
\multicolumn{5}{l}{\quad \#2 night: 2004-07-08} \\
2453195.5  & 44.9 & 11 & 0.22 $\pm$ 0.06 & 0.32 $\pm$ 0.06\\
2453195.7  & 43.5 & 50 & 0.22 $\pm$ 0.05 & 0.26 $\pm$ 0.04\\
\hline
\multicolumn{5}{l}{\quad \#3 night: 2005-05-25} \\
2453516.6  & 45.8 & 89 & 0.17 $\pm$ 0.03 & 0.25 $\pm$ 0.04\\
\hline    
\multicolumn{5}{l}{\quad \#4 night: 2005-06-23} \\
2453545.5  & 45   & 89 & 0.10 $\pm$ 0.05 & 0.18 $\pm$ 0.05\\
2453545.5  & 46.8 & 90 & 0.11 $\pm$ 0.04 & 0.22 $\pm$ 0.04\\
\hline                                   
\multicolumn{5}{l}{\quad \#5 night: 2005-06-27} \\
2453549.8  & 54   & 139& 0.13 $\pm$ 0.02 & 0.22 $\pm$ 0.03\\
\hline                                   
\multicolumn{5}{l}{\quad \#6 night: 2005-07-20} \\
2453572.5  & 58.6 & 101& 0.10 $\pm$ 0.04 & 0.21 $\pm$ 0.05\\
2453572.7  & 59.6 & 124& 0.08 $\pm$ 0.03 & 0.18 $\pm$ 0.04 \\
\hline                                   
\multicolumn{5}{l}{\quad \#7 night: 2005-08-23} \\
2453606.5  & 62.5 & 112& 0.05 $\pm$ 0.02 & 0.15 $\pm$ 0.04 \\
2453606.6  & 54.9 & 136& 0.05 $\pm$ 0.02 & 0.12 $\pm$ 0.04 \\
\hline
\end{tabular}
\end{minipage}
\end{table}

\begin{figure}
\centering
\includegraphics[width=\columnwidth]{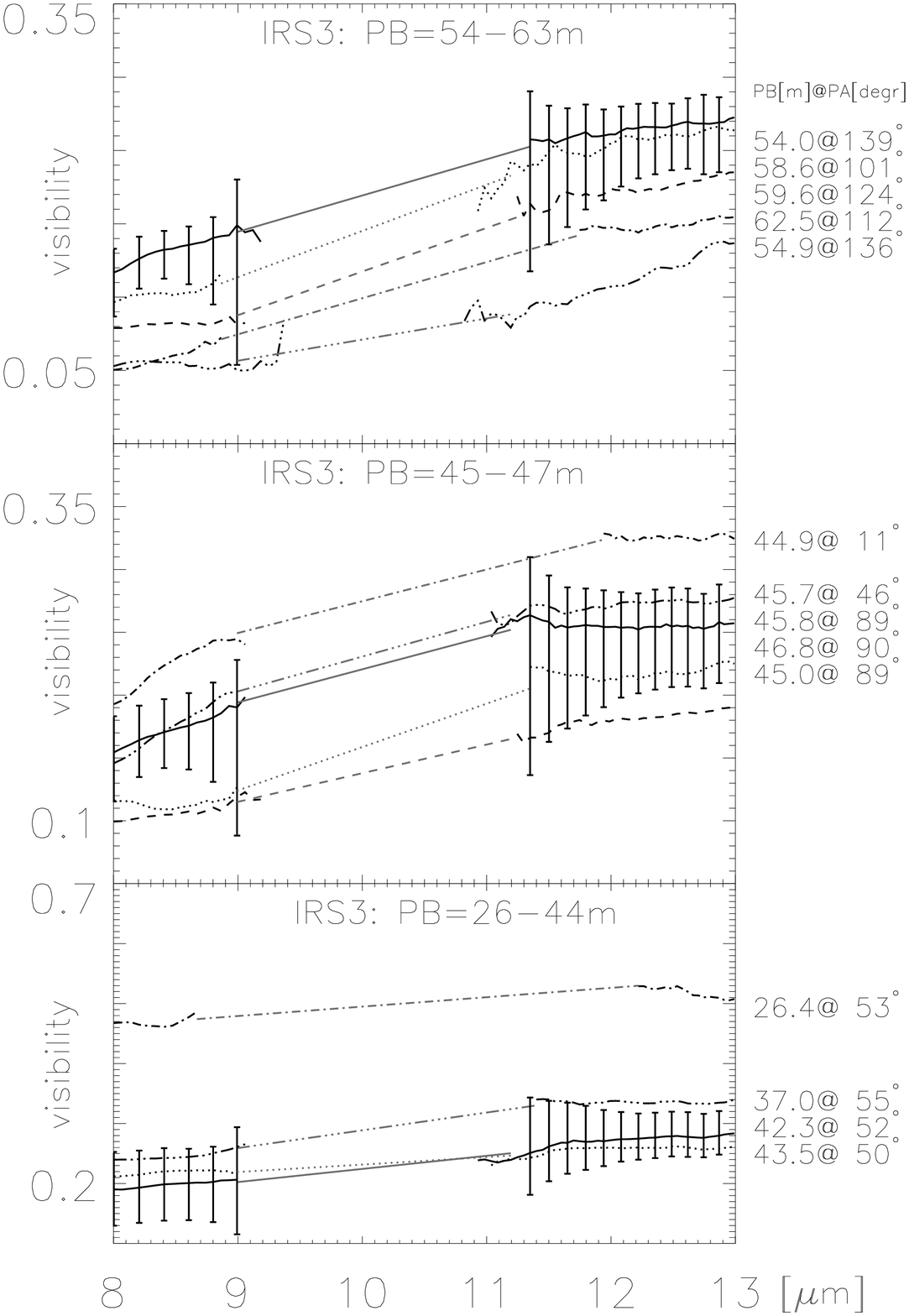}
\caption{Calibrated visibility spectra  ordered following the
  interferometric resolution to show the dependence on the projected
  baseline length (given to the right).   Each curve is plotted in a different style. For the sake of clarity we plotted only errorbars for the solid curves, and the other curves have uncertainties on a similar relative scale. The gray
  dotted lines show the used linear interpolation over the deep silicate
  absorption, where we do not have reliable data. 
 Note the different scaling
  of the panels.
}
\label{fig:31}
\end{figure}


The measured visibility moduli are given in Table~\ref{tab:3} . 
Following Sect.~\ref{sec:22} we did not find within the
uncertainties any deviation from a smooth visibility slope over the
full $N-$band. 
Since no correlated fluxes
have been measured at the center of the silicate absorption, we give the mean visibility and its accuracy for
two adjacent wavelength intervals
in Table~\ref{tab:3}.
The fully calibrated visibility spectra are shown in
Fig.~\ref{fig:31}.
Note the two outliers at the bottom of the
  upper and the middle panels. With respect to their projected baseline
  length, they show visibilities too low to be consistent with the
  other data. This is most probably an artefact due to bad beam
  overlap at those observing times.

\subsection{Probing the circular symmetry in uv-space}
\label{sec:32}

\begin{figure}
\centering
\includegraphics[width=0.8\columnwidth]{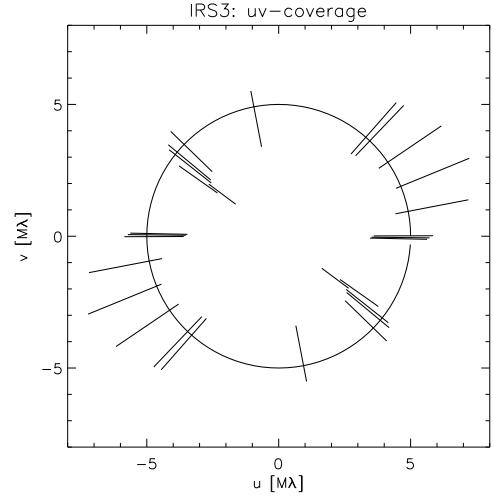}
\caption{uv-coverage of all our observations. The overplotted ring
  indicates a uv-radius of 5 M$\lambda$. We have investigated
  the degree of circular symmetry along this annulus
  (Fig.~\ref{fig:36}). North is up, East to the left.}
\label{fig:34}
\end{figure}

\begin{figure}
\centering
\includegraphics[width=\columnwidth]{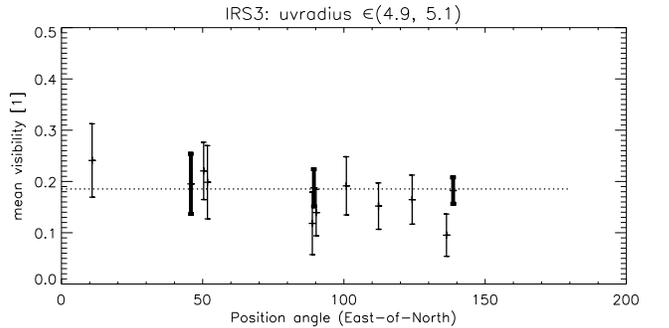}
\caption{Measured visibilities at a uv-radius of 5~M$\lambda$. To
  guide the reader's eye we have overplotted a horizontal 
  representing total circular symmetry, which was fitted to the data
  of best photometric quality in both beams (three thick
  error bars). If a linear correlation with arbitrary slope is fitted
  to the full dataset, and
  a negative slope shows a $\sim$~10\% smaller reduced $\chi^2$,
  which could indicate asymmetry or a wavelength-dependent
  size (Sect.~\ref{sec:32}).}
\label{fig:36}
\end{figure}

\begin{figure}
\centering
\includegraphics[width=\columnwidth]{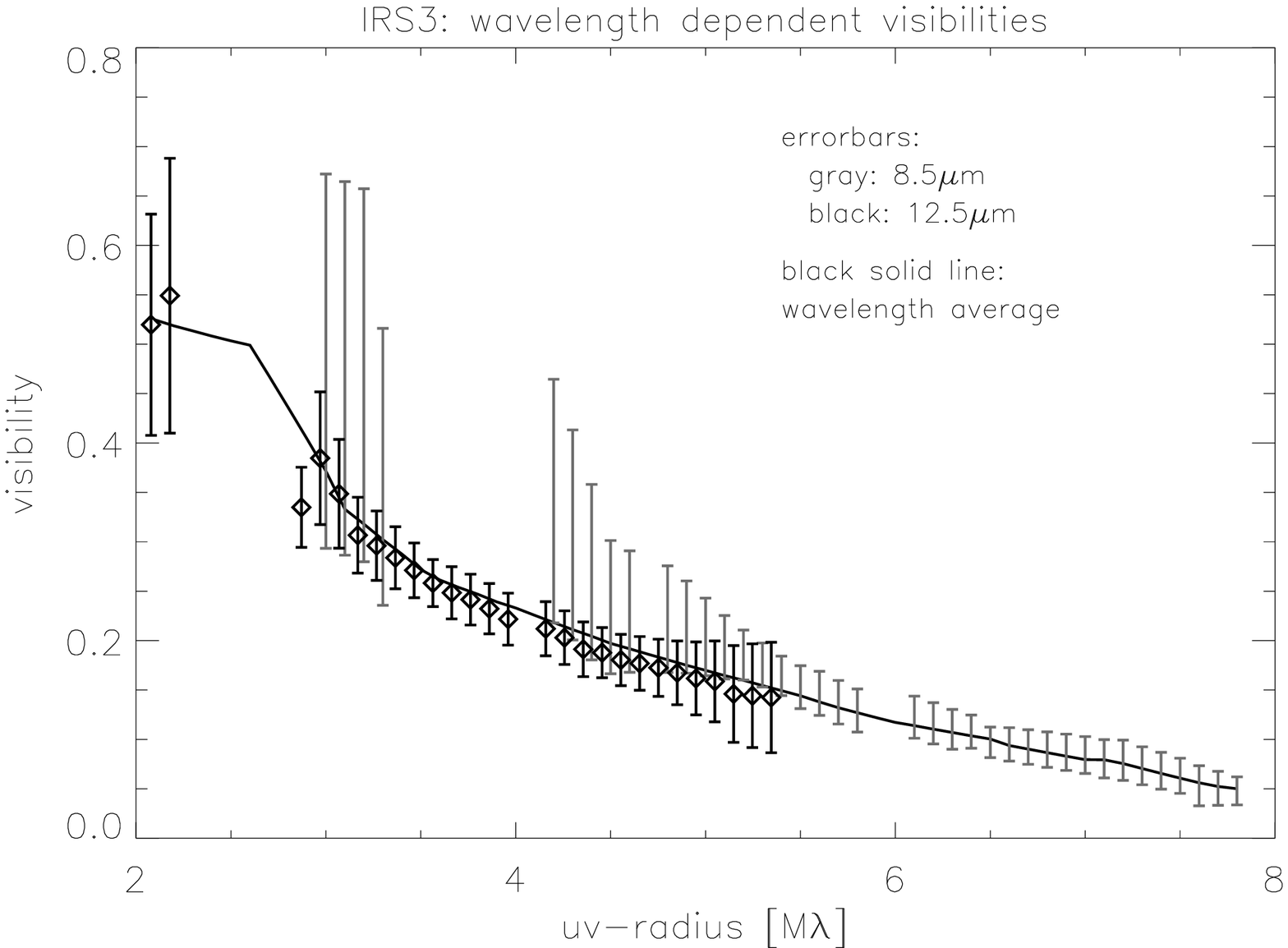}
\includegraphics[width=\columnwidth]{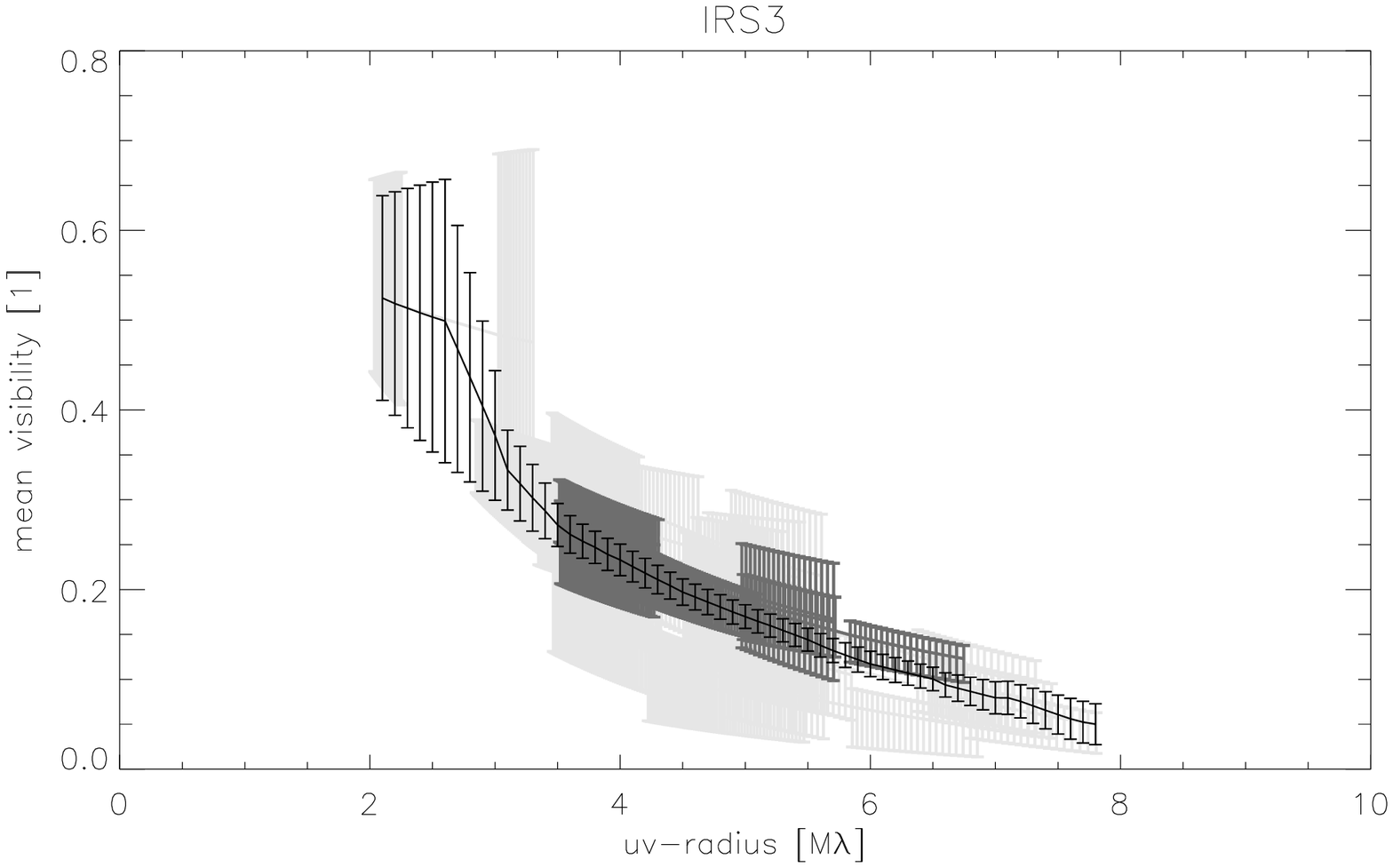}
\caption{Azimuthal data average in both panels. Upper panel: error bars represent the data taken at 8.5 (gray) and 12.5~$\mu$m (black \& diamonds), respectively. 
In the lower panel, the same spectral average (black solid
  line and error bars) is shown with 
all measured visibilities (error bars in light gray). The three datasets with the best beam quality are highlighted in dark gray. To calculate the
  average, we interpolated linearly over the silicate absorption where necessary.
}
\label{fig:37}
\end{figure}

Circular symmetry in uv-space implies circular symmetry of the
brightness distribution.
To probe the variation in the measured visibility with changing position angle (PA), we compared our
data at a  fixed uv-radius, where most data were obtained.
In Fig.~\ref{fig:34} we show the uv-coverage of the entire dataset,
overplotted with a ring of constant uv-radius (5~M$\lambda$).  
  Note that the definition of uv-radius (= projected
  baselinelength/observing wavelength) leads to radial lines in the figure and to the situation where along the circle datapoints at different PA might be observed at different wavelengths.
The uv-radius gives the angular resolution and should
show similar visibilities at different PA in the case of
circular symmetry. 
In Fig.~\ref{fig:36} the mean visibilities at $R_{\rm uv}\,=\,$5~M$\lambda$
  of each dataset are plotted.
The overplotted horizontal indicates that the full dataset still conforms
to total circular symmetry.
Although some data points do not perfectly coincide with circular
symmetry, it has to be remembered that the shown error bars of single
measurements may be underestimated, since their estimation relies on
averaging several calibrator measurements of the observing night,
leading to an average uncertainty that may be exceeded in individual cases.

On the other hand, the drawn horizontal shows that a slight deviation
from circular symmetry is possible.
The values around PA=(120$\pm 30^\circ$) appear to lie on average
below the values at smaller PA. 
Possible reasons for such circular asymmetry are discussed in
Sect.~\ref{sec:41}.

But it has to be mentioned that the data points with the best
photometric quality (used to fit the horizontal in Fig.~\ref{fig:36})
have nearly identical visibilities at different PA.
And a comparison with the uv-coverage (Fig.~\ref{fig:34}) shows that the longer
wavelength spectral channels have been considered
at those PA with a tendency towards lower visibilities (Fig.~\ref{fig:36}),
i.e. the    {possibly indicated deviation} from circular symmetry in
 Fig.~\ref{fig:36} might in fact derive from the slightly wavelength-dependent size of IRS~3. We do not expect to observe wavelength-dependent sizes due to line emission or absorption of a certain layer, since the visibility
spectrum over the $N-$band is apparently free of spectral line features
within the uncertainties.

But a slight change in the shape and the  size scale of the brightness distribution
of the dust can be expected due to  {typically} lower temperatures of larger, outer dust {layers} dominating at longer
wavelengths. This
interpretation can easily explain the deviations from perfect circular symmetry in Fig.~\ref{fig:36}  and is further backed up by a slight increase of size
with wavelength indicated by the wavelength dependent analysis of the
data (Sect.~\ref{sec:53}). 

{Circular symmetry implies a vanishing differential phase}, which remains in the data after coherent
  averaging and calibration. We did not find differential phases
  along the $N-$band spectrum larger than the remaining scatter of less than $\pm$ 5$^\circ$
  around zero. Based on the circular symmetry found, we show the PA-averaged data plotted
over the uv-radius in the lower panel of Fig.~\ref{fig:37}.
Also the black solid line in both panels is the wavelength average of all data and lies smoothly between data sets at 8.5~$\mu$ and 12.5~$\mu$m (upper panel). 
This instead indicates a slight size increase in the brightness distribution with wavelength than a qualitative change because the overall trend of the visibility with baseline length coding the brightness distribution is similar over the $N$-band.

Naturally the merged data represent an average brightness distribution at the cost of losing the slight wavelength-dependence but gaining additional and more accurate datapoints, since we can use all spectral channels together.
Note in the lower panel in Fig.~\ref{fig:37} that statistical data averaging at one uv-radius normally reduces the errorbars  with respect to the original data.
The wavelength-averaged data is used only in Sect.~\ref{sec:49} to derive an overall description of the underlying brightness distribution, which itself is used as a basis for describing the wavelength dependence of IRS~3 by fitting the average model to the wavelength-dependent data (Sect.~\ref{sec:53}).
The three datasets of highest photometric quality are
highlighted. 
This suggests that, beyond $\sim$~6~M$\lambda$, the averaged uv-data
(black solid line) indicate visibilities, which are too low since they lie below the best
data. 

\subsection{Photo-spectrometry}
\label{sec:33}

\begin{figure}
\centering
\includegraphics[width=\columnwidth]{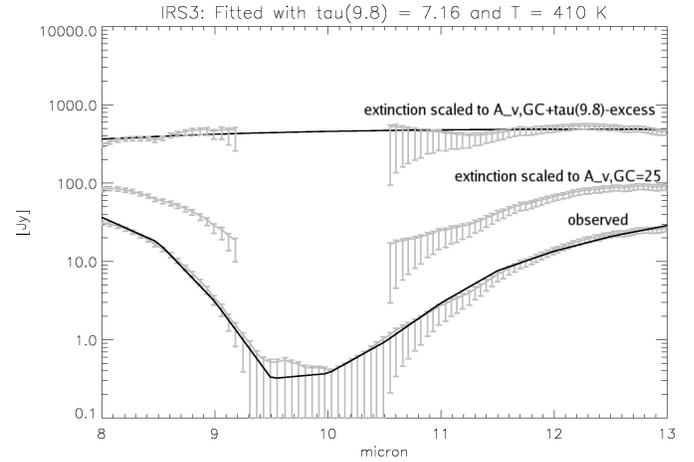}
\caption{Flux-calibrated and dereddened photometry. The upper
  spectrum is dereddened with $\tau _{9.8}=7.2$, and its error bars indicate the wavelength intervals used to fit the temperature. The middle spectrum is
  dereddened with $\tau _{9.8}=3.3$, which corresponds to the standard
  average optical extinction of $A_{\rm V}$=25 towards the central parsec, assuming
  the extinction law by \citet{2001A&A...366..106M}. 
  The lower spectrum is the extinguished, measured spectrum.
  The black solid lines show the extinguished and dereddened $\chi
  ^2$-minimized temperature fit of $T$=410~K.}
\label{fig:33}
\end{figure}

In Fig.~\ref{fig:33}, the flux-calibrated spectrum of IRS~3 is
shown based on the reduction described in Sect.~\ref{sec:24}. 
Only the data of best photometric quality and AO correction have been
taken into account. 
The three resulting data sets, observed in July~2004, May~2005, and
June~2005, do not show significant photometric variability beyond the
general uncertainties.
This agrees with the study of flux-variable sources in the GC by
\citet{1999ApJ...523..248O}, who find no variability for IRS~3.

\subsection{Morphological interpretation of the visibility data}
\label{sec:5}

\begin{figure}
\centering
\includegraphics[width=\columnwidth]{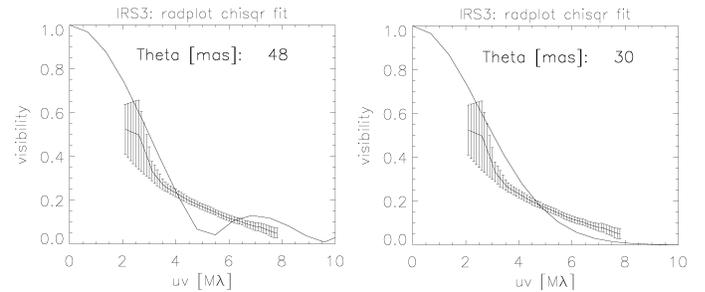}
\caption{Best-fit models of the brightness distribution with a {\em single} component
model; left panel shows a uniform disc, right panel shows a Gaussian.
The angular diameter is indicated.}
\label{fig:38}
\end{figure}
\begin{figure}
\centering
\includegraphics[width=0.9\columnwidth]{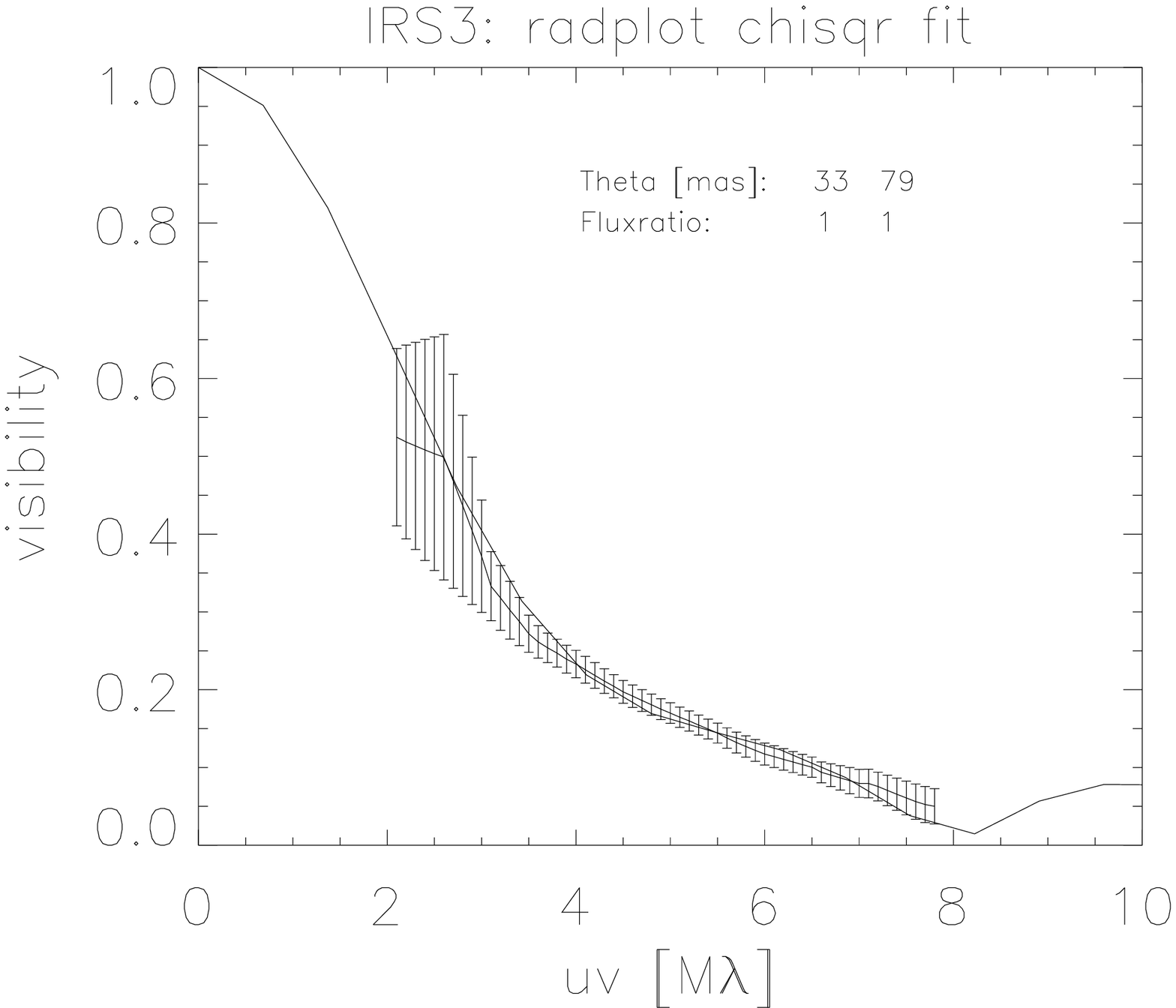}
\includegraphics[width=0.9\columnwidth]{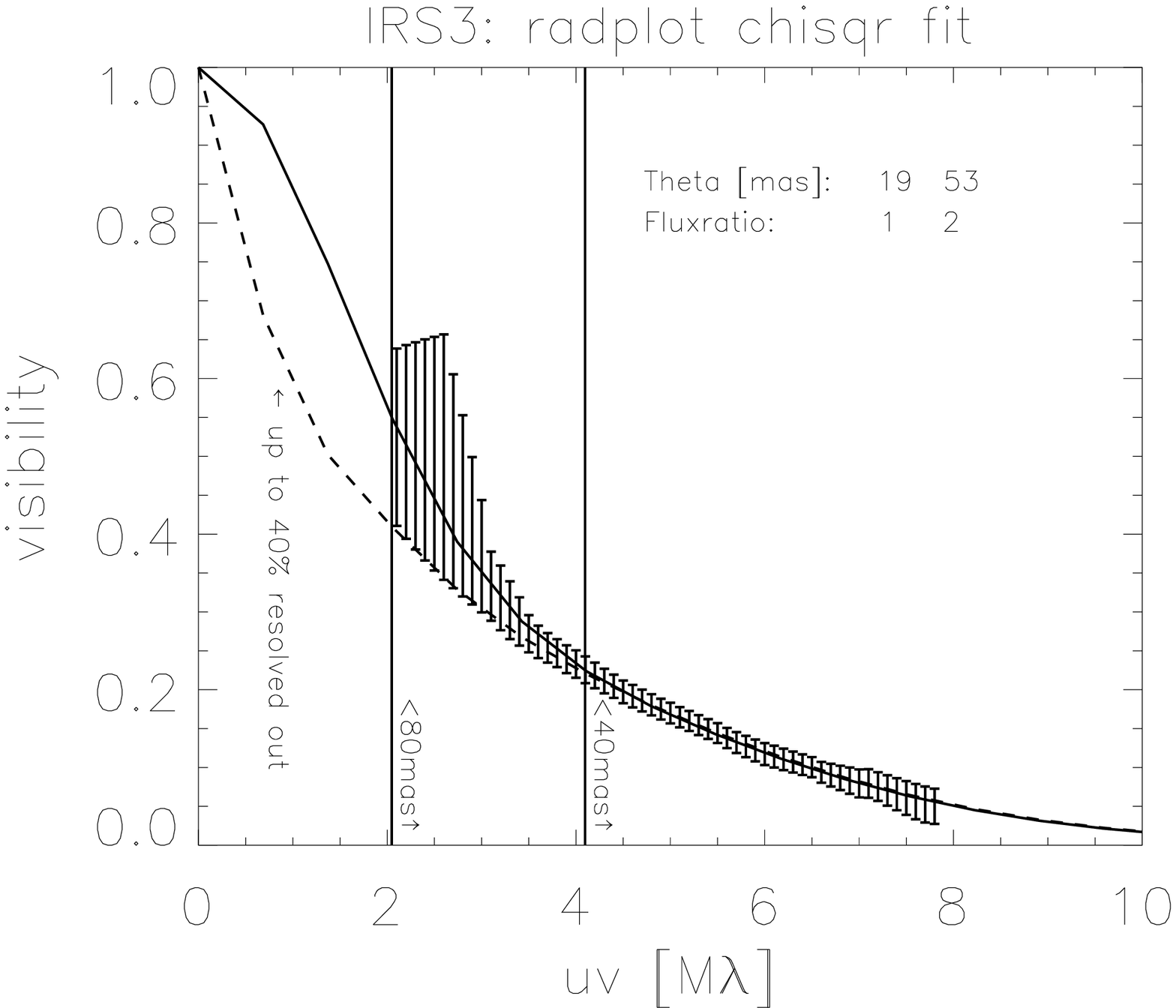}
\caption{Best-fit models of the brightness distribution with a
  {\em two}-component model. Up: two uniform disc components; Bottom: two
  Gaussian components. The error bars are the azimuthally averaged data
  (Fig.~\ref{fig:37}, Sect.~\ref{sec:49}). In addition, the probed spatial scales are indicated by vertical lines. To the right of each of these lines, the visibility of a Gaussian component of the indicated FWHM would contribute less than 10\% of its flux. The dashed line stems from a model adding a third Gaussian of arbitrary size larger than 80~mas and illustrates that up to 40\% of the total flux could have been resolved out by the interferometer.
}
\label{fig:39}
\end{figure}

In this section we explain the MIDI data by a model for the brightness distribution, which is as
simple as possible, but as complex as needed.
Although the results cannot compete with detailed radiative transfer
models, they summarize the average order-of-magnitude properties in
terms of morphological shape, size, and flux.
Such a heuristic model is therefore an important check and starting point for further
analysis, discussion, and interpretation of the data (Sect.~\ref{sec:4}).
Thanks to the large multi-wavelength data set, we get a non-trivial model of the dust distribution around IRS~3.

\subsubsection{General structure}
\label{sec:49}

Following the results of Sect.~\ref{sec:32},
we assume a circularly
symmetric brightness distribution, at first of a wavelength-independent size, {to derive the general shape of the brightness distribution}.
The full dataset is shown in a radial uv-plot in Fig.~\ref{fig:37}.
The two simplest, {but often applicable}, circularly symmetric brightness distributions are a
uniform disc and a Gaussian. 
Figure~\ref{fig:38} demonstrates that the visibility
moduli cannot be described by such a simple single component.
Despite the conservative error estimates, neither model can
reproduce the data even approximately.

In contrast, {\em two} superposed components of different
sizes and flux ratios are sufficient to model the data (Fig.~\ref{fig:39}). 
Both two uniform discs and two Gaussians
fit the
  error bars satisfactorily, but the Gaussians are closer to the measured data:  the
  reduced best-fit $\chi^2$ is about ten times smaller for the two-Gaussian model.
Furthermore, the data reduction indicates that the visibility at low uv-radii ($\sim
2.5$~M$\lambda$) is probably overestimated and should be expected to
lie in
the lower half of the indicated error bars.
This would favor a Gaussian shape for the larger component. 
{But as shown by the dashed line in the lower panel of Fig.~\ref{fig:39}, the intermediate 50~mas spatial scales are relatively loosely constrained to contribute between 20\% and 70\% of the total flux detected by MIDI. Up to 40\% of the total flux could have been resolved out by the interferometer due to the lack of shorter baseline information. 
Since with the current visibility dataset there are no means to further constrain the outer flux contributions, we stay with the two component model representing the data and outline where further analysis suggest that indeed some source flux could have been resolved out (Sect.~\ref{sec:44}). 
However, the measured correlated flux at the {\em higher} spatial frequencies (the smaller component of the model), which is most important for the further analysis and contributes about one third of the total flux, are {\em not} significantly influenced by this uncertainty.} 

At longer uv-radii the smaller component dominates. 
A closer coincidence with the data suggests a Gaussian shape for the
smaller component, too.
But the final distinction between Gaussian and disc shape  of the smaller
component requires additional data at longer baselines.
In the case of a uniform disc, the visibility would increase again around
$\sim$~10~M${\rm \lambda}$ as indicated by the overplotted model in
Fig~\ref{fig:39} (upper panel).
We investigated the uv-space $\ge$~12~M${\rm \lambda}$ with the
longest UT-baseline (UT1-4:~130m) without fringe detection.
This  might support a Gaussian shape for the smaller component. 
But it is also possible that the visibility increase, indicative of a
uniform disc shape of the inner component, was too small to be
detected with these long baselines. However, this non-detection at the long baseline supports the primary finding of our source modeling that the spatial scales of the smaller component are  also {\em resolved} by the VLTI leading to significant constraints on the physical interpretation of the data (Sect.~\ref{sec:4}). 

{At this point there is no indication that the two Gaussian
  components used to represent the data {must} stand for two
  physically distinct entities, such as two dust layers of different
  radii. 
At the moment they simply appear convenient for describing the data, and
  the larger Gaussian can be seen as representing the wings of the
  observed brightness distribution. 
In the later sections, physical parameters such as flux-calibrated spectra and derived temperatures support the idea of several physical components, but a more complicatedly shaped single structure cannot be completely excluded. 
In general the situation is similar to analyzing the uv-data of
  radio-interferometric observations \citep[e.g. of a quasar radio jet
  in ][]{2005A&A...438..785P}, where Gaussian components are used to
  represent {\em mean} properties of the observed structure, such as
  size, flux, and location. 
To avoid confusion we simply keep referring to the two Gaussians as
  being two components. This understanding is further supported by the later analysis.}

\subsubsection{Wavelength dependence}
\label{sec:53}

\begin{figure}
\centering
\includegraphics[width=.9\columnwidth]{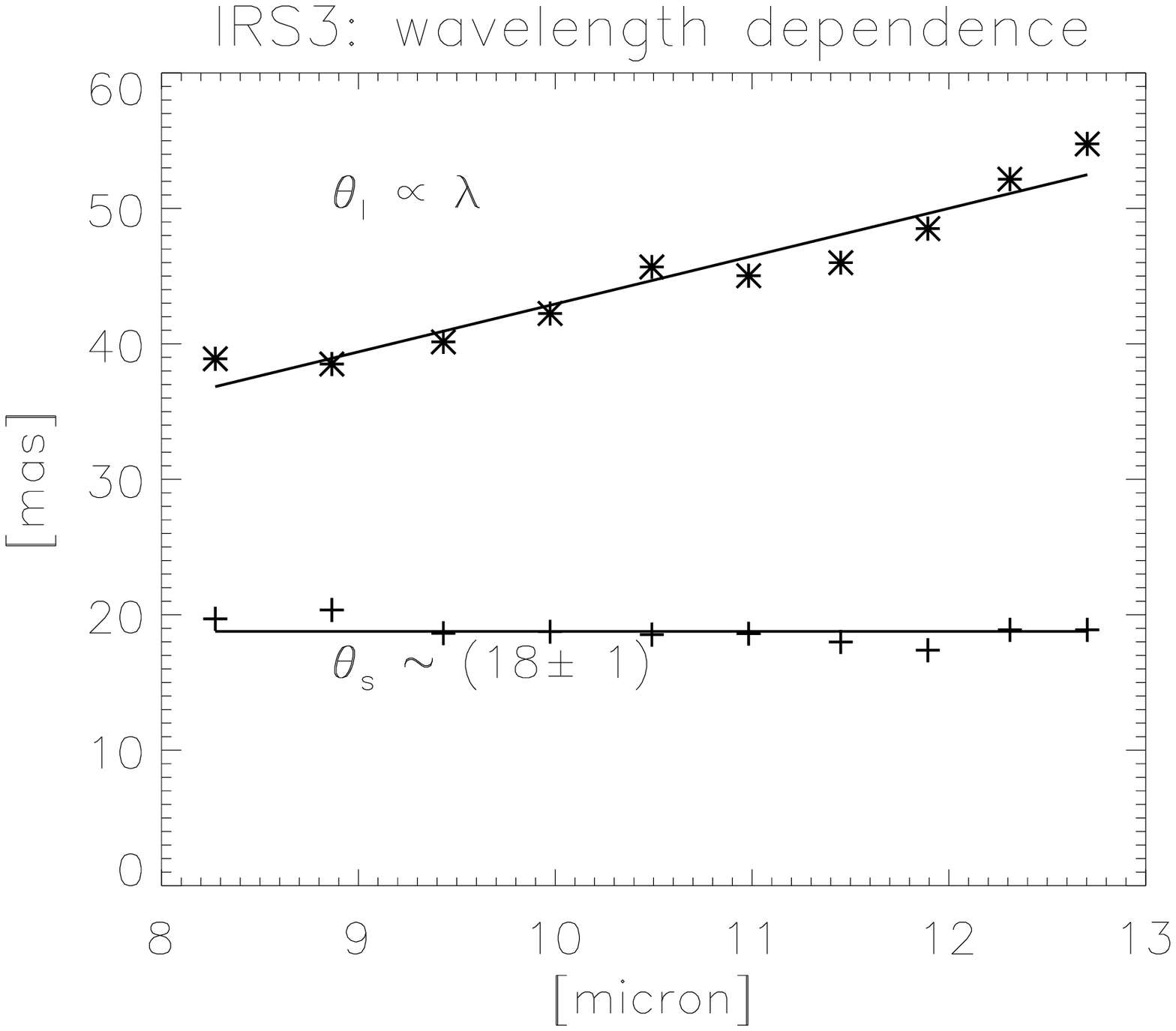}
\includegraphics[width=.9\columnwidth]{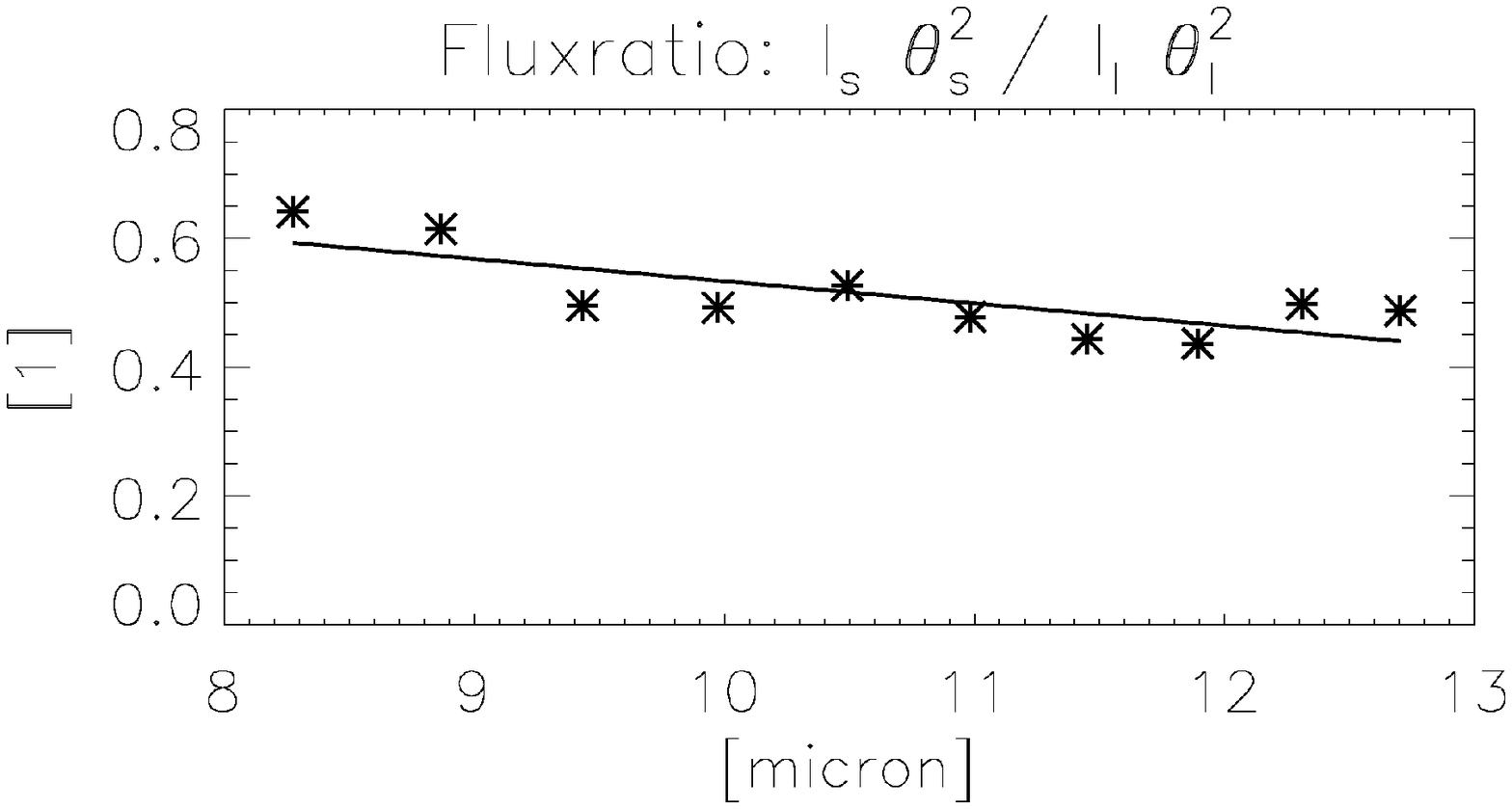}
\caption{Wavelength-dependent model consisting of two Gaussians. 
The trends are discussed in the
text. Around 10~$\mu$m, the Gaussians are fitted to the interpolated
data points. $I_{\rm s/l}$ denote the intensities (in arbitrary units)
and $\theta_{\rm s/l}$ the FWHM of the smaller and the larger
components, respectively. The flux ratio calculates as $(I_{\rm
  s}\,\theta_{\rm s}^2)/(I_{\rm l}\,\theta_{\rm l}^2)$.  In
Sect.~\ref{sec:47} the physical properties of the inner component are
analyzed by RT calculations. {The Gaussian parameters have been fitted to the directly measured visibilities in the spatial frequency domain.}}
\label{fig:53}
\end{figure}

The spectroscopically resolved MIDI data allow the investigation of
wavelength dependence of the observed brightness distribution. 
 We base this analysis on the two-Gaussian model of the previous
section, since it fits the wavelength-averaged data set perfectly. A model-independent analysis of the observed brightness distribution by Fourier-transforming the data \citep[based on the fundamental van~Cittert-Zernike theorem of
  interferometry, e.g.][]{2006iosi.book.....L} confirms of our model approach, but the limited spatial
  frequency coverage hampers further {\em model-independent} conclusions from such an analysis.
We describe the existing wavelength-dependence of the data in terms of changing
  sizes and relative flux contributions of the two Gaussians, keeping the comment at the end
  of the previous section in mind. 
Furthermore, the change of the fit parameters with wavelength turns out to be
  reasonably smooth (Fig.~\ref{fig:53}). This finding backs the application of the wavelength-independent model of Sect.~\ref{sec:49} as a basis for fitting the wavelength dependence.

We fit intensity and size of two superposed Gaussians to the data
(Fig.~\ref{fig:53}).
We bin the data to a 0.5~$\mu$m sampling, using error-weighted
visibility averages. 
The unreliable data around the center of the silicate absorption are
interpolated using a $\chi^2$-fit of a quadratic curve to the spectral channels of good
SNR. 
In Fig.~\ref{fig:31} the individual regions of reliable data are indicated by
  error bars. The interpolated data are used in the central
  wavelength interval only, which is devoid of error bars.
 The respective dependence of the
  visibilities on the
  position angle is more visible in Fig.~\ref{fig:36}. 

In Fig.~\ref{fig:53} the resulting best-fit parameters are shown.
Using $\chi^2$-minimization, the uncertainties of the fit are nearly the same as the {scatter of the data} around the overplotted linear correlations.
A slightly larger systematic uncertainty might be introduced by the
estimated errors of the individual data points.
To address this, we re-fit the data with increased weighting of the
three datasets of best photometric quality. 
The overall trends are similar, but the size of the larger component in the two-component model approach may be
underestimated in Fig.~\ref{fig:53} by 5-10~mas, {which indicates more flux on low spatial frequencies}. The
  FWHM of the smaller component, $\theta_s$, may
show a slight size-increase with wavelength by about 3~mas, and the
flux ratio increases towards the smaller component ($F_{\rm s}/F_{\rm
  l} \sim 0.7$).
This increased flux ratio can be understood by lower
photometric quality, typically decreasing the average visibility due to
an imperfect beam overlap.
But a decreased visibility means increased {\em relative} brightness of the larger component. 
Thus, probably a few of the data sets of lower photometric quality systematically show visibilities that are  too small, {thus
  artificially implying} structures that are less concentrated than the real ones
  (see the bottom curves of the upper and central panel of Fig.~\ref{fig:31}, which {\em apparently} do not follow the visibility trend of the other curves with respect to the baseline lengths).

The most intriguing result is that the smaller component ($\theta_s$)
shows a roughly constant
FWHM of about 18~mas, while the larger
shows a significant linear size increase  with wavelength. 
{This {might be} an indication that we directly resolve the inner
  zone of dust formation at all
  wavelengths, which has a fixed size, and that the dust
  shell might be carbon-rich \citep[][]{1996MNRAS.279.1011I}}.

Furthermore, the relative flux contribution of the larger component 
increases with wavelength, suggesting that the larger component {represents the}
outer, cooler dust around the central object. {Again this is a more qualitative statement. The second component stands for the more extended flux of IRS~3, its properties average over the conditions at the outer parts, but a smooth transition between the inner and outer regions around the star cannot be excluded.}
Temperatures and luminosities {of dust on the size scales of} both components are derived in Sect.~\ref{sec:44} based purely on the
interferometric data.

\section{Discussion}
\label{sec:4}
\subsection{Interstellar absorption and the composition of the circumstellar absorbing dust}
\label{sec:45}

\begin{figure*}
\centering
\includegraphics[width=\textwidth]{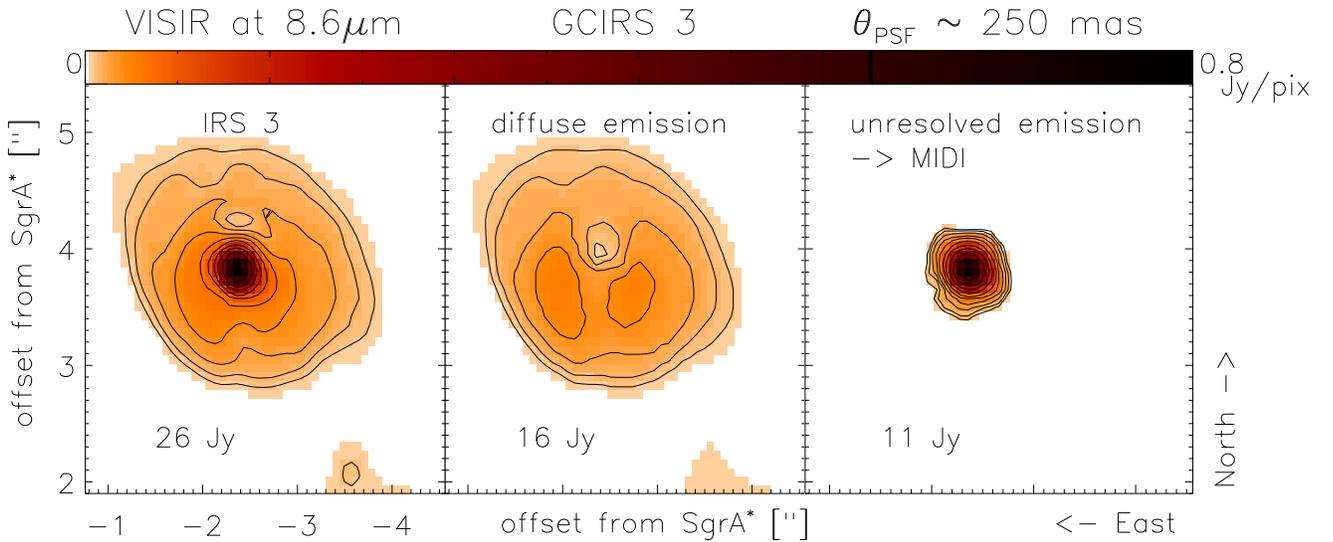}
\caption{Our VISIR 8.6~$\mu$m imaging data of a  4'' field of view
  centered on IRS~3. Left: A zoom of Fig.~\ref{fig:62}. Middle: The
  diffuse emission with subtracted PSF. Right: The flux-calibrated
  PSF, i.e. the unresolved emission seen by MIDI. The data reduction is detailed in
  \citep{2007A&A...462L...1S}. All images have the same color coding as indicated
  at the top and the same contour lines for ease of comparison. The
  logarithmic contours levels are 1.6$^{n}\cdot$7~mJy. The {integrated} flux of the
  different components is given in the figures. Note that the diffuse
  flux peaks below 40~mJy/pix. 
}
\label{fig:60}
\end{figure*}

\begin{figure}
\centering
\includegraphics[width=0.8\columnwidth]{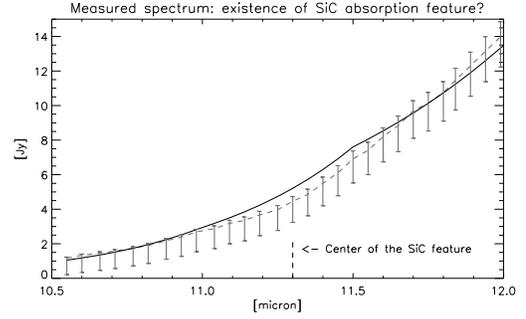}
\caption{Zoom of the lower spectrum in Fig.~\ref{fig:33} into the wavelength interval of
  the 11.3~$\mu$m feature of SiC. The measured data is plotted in
  (dashed) gray, and the black solid line represents the best-fit extinguished
  blackbody SED of $T\,=\,410$~K.  }
\label{fig:15}
\end{figure}

Following the most recent published results, we assume a spectral profile of
the interstellar absorption towards the GC, as published by
\citet{2001A&A...366..106M}, and an average visual extinction of $A_{\rm V} =
25$ towards the GC
\citep{2003ApJ...594..294S,2005A&A...433..117V}. \citet{2001A&A...366..106M} incorporated
into their model that the mean interstellar dust towards
{sources in the central 2~pc of the Galaxy} shows a relatively stronger silicate absorption than in
the solar neighborhood \citep{1985MNRAS.215..425R}. 

In the NIR no strong local increase in the interstellar extinction towards
the region of IRS~3 is found \citep[e.g. most recently confirmed
  by][]{2007A&A...469..125S}.
However, several authors claim {additional} silicate absorption in the $N$-band only along the
line of sight to IRS~3, possibly intrinsic to the source
\citep[e.g.][]{1985MNRAS.215..425R,2006ApJ...642..861V}.

{A first glimpse of the probable location of this additional
  extinction is given by our high-resolution single-telescope $N$-band
  imaging with the new VLT/VISIR instrument. 
In Fig.~\ref{fig:60} the
  complete emission of IRS~3 (left panel), unresolved by earlier imaging, is
  clearly resolved into a diffuse  and a compact component (central and right panels, respectively). 
Although the
  total diffuse flux is even larger than the compact flux, its surface
  brightness is very low and hidden in the noise of all MIDI
  data due to the shorter integration times. 
Thus most of the diffuse flux is not included in the single-telescope MIDI photometry ($F_{\rm
  T,\,\lambda}$), although it is observed
  at similar spatial resolution to the VISIR data. 
Therefore the flux-calibrated MIDI $F_{\rm T}$ spectrum (Fig.~\ref{fig:33}) fits the unresolved flux shown in the right panel. 
The two components, fitted to the interferometric data
  (including the possible fraction of fully resolved flux; Sect.~\ref{sec:5}) and discussed in the following sections,
  together make up the unresolved emission in the right panel of Fig.~\ref{fig:60}, since the single-telescope observations provide an angular resolution of about 250~mas.
  To clarify the situation, we speak of {\em local, interstellar} silicate
  absorption in addition to the GC average, if the absorbing silicate
  is located in the dust, which radiates the diffuse
  emission that is directly visible only in the VISIR data. In contrast, absorption in the inner dust components,
  resolved by the MIDI flux and visibility estimates, is labeled as {\em intrinsic, circumstellar} absorption.}

Our MIDI spectro-photometry  confirms the existence of a broad 9.8~$\mu$m silicate absorption feature
remaining  in the data after correction for standard GC extinction.
We show the measured spectrum in Fig.~\ref{fig:33} (lower curve).
For the dereddening of the spectrum, we used the $\mu$Cep emission profile of
the silicate feature, as realized in the extinction law by
\citet{2001A&A...366..106M} for lines of sight to the GC. 
Several authors state that this profile matches both
the local ISM absorption and the GC interstellar silicate absorption
profile best
\citep[although at different relative optical depths;][]{1984MNRAS.208..481R,1985MNRAS.215..425R,2006ApJ...637..774C}.

Our IRS~3 spectrum of a spectral resolution of $R=30$ shows this
coincidence perfectly. 
We reddened a single blackbody of variable temperature with an
absorption spectrum of the normalized shape of the silicate feature
seen in emission towards $\mu$Cep \citep[see][]{2001A&A...366..106M}
and a variable optical depth $\tau_{9.8}$. 
The shapes of the  observed and dereddened spectra coincide with the
spectra shown by \citet{1985MNRAS.215..425R} (Fig.~\ref{fig:33}). 
The best $\chi ^2$-fitted parameters are a blackbody temperature of $T=(410\pm 30)$
and $\tau_{9.8} = (7 \pm 0.5)$.
This temperature resembles other $N-$band measurements \citep[e.g.][found a
color-temperature of about 400~K]{1985ApJ...299.1007G}, but is below the 800~K derived from
 $K-$
and $L-$band data \citep{2004A&A...425..529M}. 
This indicates that a single blackbody may not be appropriate for describing the
complete NIR-MIR SED. 
Although the hotter component, dominating the $K-$ and $L-$band, cannot be
resolved against the cooler outer dust shell by the single telescope
spectrum or image at 10~$\mu$m with additional SED information, our interferometric 10~$\mu$m data alone can resolve it (Sect.~\ref{sec:44}).

Since the center of the absorption feature is hidden in the noise of
the background subtraction, a certain level of uncertainty remains
in the estimation of $\tau_{9.8}$, but the spectrophotometric quality
of the wings is good enough to exclude $\tau_{9.8} \le 6.5$. 
Assuming $A_{\rm V}=25$, this means $A_{\rm V}/\tau_{9.8}\le 4$.
This is a remarkable result, since it doubles the silicate MIR
optical depth towards IRS~3 with respect to the average of the GC
region \citep[$(A_{\rm V}/\tau_{9.8})_{\rm GC} \approx$
  8-10;][]{1985MNRAS.215..425R}, which itself is twice as deep as
$A_{\rm V}/\tau_{9.8}$ in the solar neighborhood.
This is shown by the middle spectrum in Fig.~\ref{fig:33}, which
is the measured spectrum corrected for standard GC values of extinction
($A_{\rm V}=25$ and $(A_{\rm V}/\tau_{9.8})_{\rm GC}$). The remaining
  silicate absorption is obvious.
The aforementioned authors quantify for the first time an extra
$\tau_{9.8}\approx$ 0.8 for IRS~3 in addition to already enhanced the GC-average.
Although the spectral resolution of both datasets is comparable, the
spatial resolution of the MIDI photometry data\footnote{which is the
  VLT 10~$\mu$m resolution of $\sim 250$~mas, not the interferometric resolution} is increased by at least an order
of magnitude.

In addition, at 11.3~$\mu$m a significant drop in the data below the fit
is suggested (Fig.~\ref{fig:15}), even on the logarithmic scale shown in Fig.~\ref{fig:33}. 
This further absorption feature, {in addition to the dominating broad
silicate absorption}, can be attributed to SiC,
which  peaks around 11.3~$\mu$m and has a much narrower spectral width than the
interstellar silicate feature. 
If we exclude the data from the wavelength region around 11.3~$\mu$m for the
$\chi^2$ minimization, the discrepancy
between data and fit around the center of the SiC feature  
becomes even stronger, although the fitted temperature
and $\tau_{9.8}$ remain constant in the given interval of uncertainties. This
further supports the existence of a SiC absorption feature towards
IRS~3, but the sampling of the applied extinction law, which does not contain the SiC feature, is with
0.5~$\mu$m at MIR wavelengths not high enough to properly sample the SiC
feature. The kink of the blackbody fit at 11.5~$\mu$m is probably an artificial feature. Thus a definitive answer regarding the existence of absorbing SiC cannot be given.

Arguments against an intrinsic silicate
absorption in the immediate circumstellar dusty environment of IRS~3, which
could be evoked by a deep O-rich dust shell, are:
\begin{itemize}
\item Most silicate-rich dust shells show the silicate feature in
  emission. Similarly the SiC feature is typically found in emission
  in the dust shells of evolved stars. Since our
  estimated optical depth $\tau_{9.8}$ is already very deep, an
  even larger amount of absorbing dust would be necessary to
  overcome the  circumstellar emission and result in such strong
  features as observed.
\item The spectral shape of the observed silicate absorption coincides perfectly  with the interstellar absorption features. No indications
  of circumstellar crystalline silicates are found, although
  spectroscopic data with higher SNR and spectral resolution covering the full $N-$band are needed to investigate the spectral shape in more detail.
\end{itemize}

Thus a circumstellar dust shell free of a significant amount of
 silicates appears to be a reasonable assumption
for the immediate environment of IRS~3.
This is confirmed by our visibility data, which do not show any spectral feature coinciding with
the broad shape of the 9.8-silicate feature  at any baseline length.
 
{Such a lack of an intrinsic, circumstellar silicate-rich dust
  shell and the deep interstellar silicate absorption would} favor the bright IRS~3 to be the primary target for estimating
the true spectral shape of the interstellar absorption in the $N-$band towards
central GC sources at the high spatial and spectral resolution now
available at ground-based 8~m class telescopes.

\subsection{Dust temperatures from the spectral energy distribution}
\label{sec:46}

\begin{figure}
\centering
\includegraphics[width=\columnwidth]{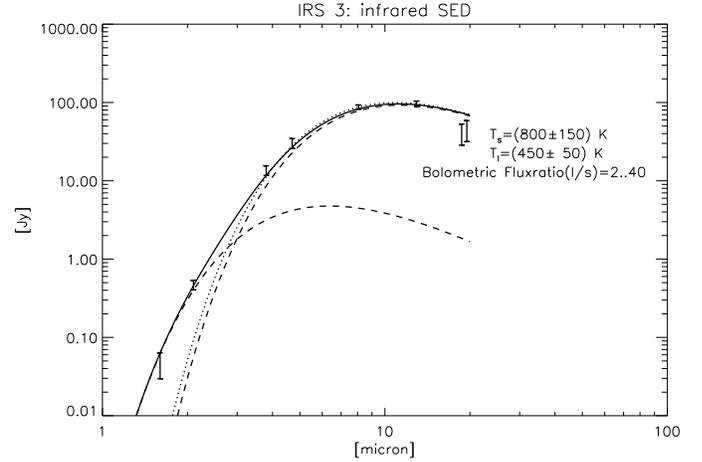}
\caption{SED temperature fit (solid line) as the superposition of two spatially unresolved blackbody
  SED (dashed lines). The dotted line close to
  the cooler and presumably larger $T_{\rm l}$ component shows a
  blackbody SED of $T\,=410$~K, which was fitted to the $N$-band spectrum
  alone (Sect.~\ref{sec:45}). The possible bolometric
  flux ratio range is given in the plot. Data are taken from
\citet[][$H-$, $K-$, $L-$, $M-$bands]{2005A&A...433..117V} and \citet[][$Q-$band]{2006ApJ...642..861V} and corrected (see text). The $N-$band
data was taken from our MIDI observations at 8 and 13 micron, outside
the silicate absorption feature. The data was extinction-corrected for
$A_{\rm V}$=25. 
}
\label{fig:45}
\end{figure}

The MIR regime is dominated by thermal dust emission. 
We have shown
in the previous section that the $N-$band spectrum of IRS~3 can be
fitted convincingly by a single reddened blackbody.
But published studies of stellar dust shells show that the
full infrared photometric information is required to describe the
optical and physical properties of the shells. 
Accordingly, we investigated the complete infrared wavelength range from 1.6-20~$\mu$m,
available at a spatial resolution sufficient for distinguishing IRS~3 from
other sources.
The corresponding dereddened SED is shown in
Fig.~\ref{fig:45}. 
The bolometric flux ratio indicated in the figure is better
  constrained by the MIDI visibility estimates (Sect.~\ref{sec:44}).

A second star close to IRS~3 has recently been classified as a Wolf-Rayet star
\citep[][]{2006ApJ...643.1011P}. 
{It is unresolved in the published medium-resolution NACO data
\citep{2005A&A...433..117V}, but we confirm this secondary star on
high-resolution NACO images, showing about 30\% and less than 10\% of
the IRS~3 $H$- and $K$-band fluxes, respectively. For our SED fit in
Fig.~\ref{fig:45} we used the accordingly reduced published $H$ magnitude.} 
Since the companion is located about 120~mas east of IRS~3, its MIR flux
could also contribute to the MIDI data.
But a significant contribution should show up as a binary pattern in
the visibility data (Sect.~\ref{sec:41}), which was not observed. 
Thus we assume negligible contamination of the flux of
IRS~3 longward of 2~$\mu$m by the WR-star, which is further supported by
  the bluer NIR SED of the secondary {as measured with NACO and expected for a hot WR}. 

To account for the uncertainties in the amount of interstellar
silicate absorption towards IRS~3 (Sect.~\ref{sec:45}), we used only the MIDI fluxes at 8 and 13~$\mu$m outside the 9.8~$\mu$m absorption
feature for the
MIR-SED.  
All data were dereddened with the Moneti extinction law scaled to
$A_{\rm V}$=25. 
As for the $N$-band, a broad interstellar silicate
absorption feature is located in the  $Q-$band.
At 20~$\mu$m, only narrow-band photometry inside the silicate
absorption was available. 
We dereddened this $Q$-band data by {an optical depth of $\tau _Q\sim 3.5$, which is derived from scaling the interstellar extinction law to fit our measured
enhanced $\tau_{9.8}$} towards IRS~3, exceeding the average GC values. {At least
  the
  plotted SED data for $\lambda \leq 13\,\mu$m should be free of any
  significant flux contribution of the diffuse VISIR component, because of
  its low surface brightness (Sect.~\ref{sec:45}) and presumably cool temperature. Only the
  VISIR $Q$-band data may contain a fraction of the diffuse flux; although
 similar to the MIDI spectra, these $Q$-band data do not reach
 the low noise level of the 8.6~$\mu$m VISIR imaging data.}

We successfully fit the extinction-corrected SED with {\em two}
blackbody spectra. 
The lower temperature, which we attribute to the outer component
($T_l=(440\pm 50)$~K), coincides with the single blackbody temperature
fitted to the $N-$band spectrum alone (Sect.\ref{sec:45}). 
This single blackbody SED is overplotted in Fig.~\ref{fig:45}, which shows immediately that for $\lambda \le
5\mu$m additional flux by hotter dust is required to explain
the measured infrared SED.
This is consistent with the findings of \citet{2004A&A...425..529M}, who fitted a blackbody temperature of 800~K to their 2-4~$\mu$m data. {That hotter dust emission {\em appears} at shorter wavelengths limits the optical depth of the outer cooler dust, assuming that the hotter dust is located inside the cooler dust (see also Sect.~\ref{sec:44}).}
The calibration of the $Q-$band emission, apparently too faint to fit the model,
was very difficult and the deviations can be fully attributed to calibration
errors (Viehmann, priv. communication) including the uncertain amount
of interstellar silicate extinction in that band.

We can state, summarizing these considerations, that no further blackbody component is needed to model the full {near- and mid-}infrared SED. 
In particular this excludes any significant stellar contribution to the
NIR fluxes of the SED, emerging from IRS~3 itself or a second star
inside the PSF.
This leads to two possible interpretations: either the inner and hotter
dust component is optically thick at NIR wavelengths avoiding any direct detection of stellar
light or the enshrouded star has a very hot continuum emission.
The absence of any stellar photospheric lines in NIR spectra of IRS~3
\citep{2006ApJ...641..891T} supports the optical thickness of the
circumstellar dust at these wavelengths.

\subsection{Dust temperatures from the interferometric data}
\label{sec:44}

\begin{table}
\begin{minipage}[t]{\columnwidth}
\caption{Temperatures from the $N-$band MIDI data based on size and relative flux estimates of the two-Gaussian model and $\tau = 0.5$ of the outer dust.  
}
\label{tab:44}
\centering         
\renewcommand{\footnoterule}{}  
\begin{tabular}{rcccc}
\hline\hline       
 Properties               &\multicolumn{2}{c}{Inner (20~mas)}&\multicolumn{2}{c}{Outer (50~mas)}\\
                &8~$\mu$m & 13~$\mu$m &8~$\mu$m & 13~$\mu$m\\
\hline
$F\,\footnote{\label{foot:36}The individual component fluxes and sizes have uncertainties of about 15\% and 10\%, respectively, within the model. The additional uncertainty range of the flux contribution of the larger component due to the lack of data at short spatial frequencies (Sect.~\ref{sec:49}) is indicated with dots.}\,[{\rm Jy}]$ & 35 & 32 & 21..55&  25..66\\
$\theta\,\textsuperscript{\ref{foot:36}}\,[{\rm mas}]$  &18 &18 &40 & 55\\
$T_{\rm Fratio}\,[{\rm K}]$ & \multicolumn{2}{c}{(610$\pm$180)}& \multicolumn{2}{c}{(460$\pm$100)}\\
$T_{\rm sphere,~8}\,[{\rm K}]$ & \multicolumn{2}{l}{(830$\pm$120)\hspace{1.2cm}}&\multicolumn{2}{l}{(480..630$\pm$70)\hspace{1.2cm}}\\
$T_{\rm sphere,~13}\,[{\rm K}]$ & \multicolumn{2}{r}{\hspace{1.2cm}(1020$\pm$210)}&\multicolumn{2}{r}{\hspace{1.2cm}(400..580$\pm$90)}\\
\hline             
\end{tabular}
\end{minipage}
\end{table}

In contrast to the previous section, here we present a derivation of
temperatures from the {\em spatially resolved} MIDI 
observations in MIR as a further step in interpreting this (and similar) interferometric data.

The simplest morphological interpretation of the data consists of two
circular symmetric Gaussian components enshrouding the same central
object (Sect.~\ref{sec:5}). One-dimensional radiative transfer calculations confirm bell-shaped brightness distributions of circumstellar dust shells
\citep{1996MNRAS.279.1011I}.
That we can {distinguish at least two} dust components suggests the outer one is optically thin {in the respective wavelengths range} and physically separated.
That is, the observed total
flux constitutes of the sum of the flux of both components{, adjusted by the optical depth $\tau$}.
From the observed total fluxes and flux ratios, we calculate the component
fluxes as
\begin{eqnarray}
\label{equ:5}
  F^{\rm d}_{\rm tot,\lambda} &=& F_{\rm s,\lambda} +  F_{\rm l,\lambda}=e^{-\tau_\lambda}\,F'_{\rm s,\lambda} + (1-e^{-\tau_\lambda})\,F'_{\rm l,\lambda} \nonumber \\
&=& (R_\lambda+1)\,F_{\rm l,\lambda}
\end{eqnarray}
where $F^{\rm d}_{\rm tot}$ is the total dereddened flux (Sect.~\ref{sec:45}),
$R$ the flux ratio between the inner and the outer dust shells as observed, {and $F'_{\rm s,l}$ are the intrinsic flux densities corrected for the optical depth. }
All measurable quantities in Eq.~\ref{equ:5} are supposed to be wavelength-dependent.
In other sections we present strong indications that the absorbing
silicate is not located in the inner circumstellar dust
(Sect.~\ref{sec:46}~\&~\ref{sec:47}).
Nevertheless, here we confine the calculation to the edges of the $N-$band
outside the silicate feature to minimize the possible corruption of the results
by faulty correction for the interstellar extinction.

In Table~\ref{tab:44} the component fluxes and FWHM-sizes for the
two-Gaussian model of Sect.~\ref{sec:5} and the temperatures derived here are given:
$T_{\rm sphere}$ is the {brightness} temperature of a spherical blackbody of radius
$\theta /2$ at GC distance emitting the observed flux, and
$T_{\rm Fratio}$ the blackbody {color} temperature derived solely from the
8-13~$\mu$m color of each component. {We corrected both
  component fluxes for an optical depth of the outer shell of $\tau
  = 0.5$. Since typically the optical depths at 8 and 13~$\mu$m
  are comparable, the color temperature does not change with $\tau$.}

{Furthermore, RT calculations of spherical dust shells as used in Sect.~\ref{sec:47} indicate that a black body spectrum of $T_{\rm Fratio}$ often convincingly fits the MIR total spectrum of a dust embedded star, particularly well for carbon-rich dust shells. But the respective $T_{\rm sphere}$ of such models, calculated again with a Gaussian brightness distribution approximation, is often significantly higher than $T_{\rm Fratio}$ of the same RT model. This is in line with the findings presented in Table~\ref{tab:44} showing that the temperatures derived here only give general trends and temperature ranges without imposing tight constraints on the physical temperature.}

We find:
\begin{itemize}
\item A reasonable increase in $T_{\rm Fratio}$ of the inner component with
  respect to the outer one. This coincides perfectly {with the analysis of the infrared SED in the previous section}. 
The MIR-interferometric data alone show the inner, hotter dust component in contrast to spatially unresolved $N-$band
  photometry. But the physical properties of the inner dust are better
  confined by the complete IR-SED and by more
  detailed radiative transfer calculations (Sect.\ref{sec:47})
\item {Physically reasonable temperatures can only be obtained with an optical depth of $\tau_{8/13}\sim 0.5$ due to the outer dust {\rm in addition} to the known interstellar extinction $A_{\rm V} = 25$. This backs the large $\tau_{9.8}$ discussed earlier found towards IRS~3 but suggests at the same time that this additional silicate absorption does {\em not} occur in the innermost circumstellar dust.}
\item {Constraining the optical depth is also important for deriving the intrinsic luminosity of $\sim 5\cdot 10^4~L_\odot$ only from the inner component fluxes. This luminosity is slightly higher than earlier estimates \citep{1978ApJ...219..121B} that were based solely on spatially unresolved $N$-band photometry. Thus it can now be excluded that any additional source outside the central 20~mas significantly contributes to the flux and heating of the dust around IRS~3, a finding confirmed in the next section. }
\end{itemize}

\subsection{Circular symmetry}
\label{sec:41}

\begin{figure}
\centering
\includegraphics[width=\columnwidth]{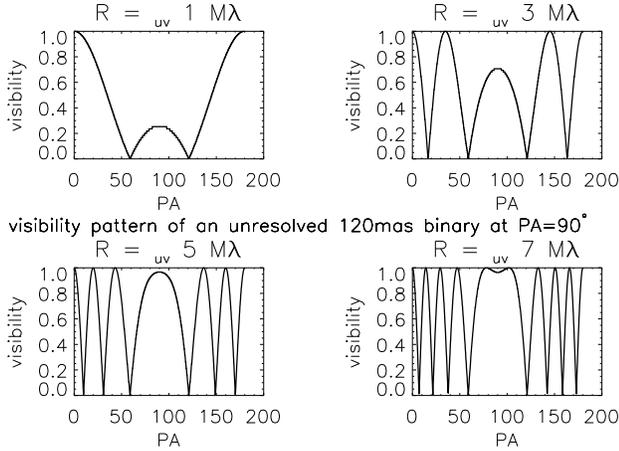}
\caption{Visibility patterns for a 120~mas east-west binary at
  several uv-radii. The individual components are not resolved.}
\label{fig:41}
\end{figure}

In their spectroscopic survey of
the central cluster, \citet{2006ApJ...643.1011P} classify the star IRS~3E, only 120~mas east of IRS~3, as a
Wolf-Rayet star of type WC~5/6. 
Since Wolf-Rayet stars of this spectral type are known to be (often
strong) dust formers, it is possible that IRS~3E is still visible in
the $N-$band.
Additionally its spectral classification as a carbon-rich WC-star does not
conflict with our interpretation of the spectral data in terms of a lack
of silicate emission.

The visibility pattern of such a binary system with 120~mas separation
and east-west orientation is shown in Fig.~\ref{fig:41} for several
$R_{\rm uv}$. 
The shown pattern has been calculated for two stars of equal
brightness, and both are individually unresolved by the interferometer.
IRS~3 {\em is} resolved by MIDI, which would decrease the amplitude of the
variation shown in Fig.~\ref{fig:41}, but 
the variation per $PA$ is defined by the binary separation only;
i.e., the imprint of a 120~mas binary system should show several
ripples over 180$^\circ$ rotation, which cannot be confirmed with our
data (Fig.~\ref{fig:36}).
Furthermore, we did not observe neither photometric
variability beyond the calibration uncertainties (Sect.~\ref{sec:33})
{or any non-zero differential phases (Sect.~\ref{sec:32})}.

The simplest deviation from circular symmetry is an elliptical morphology or, more generally, a brightness distribution of different apparent
extension in orthogonal directions.
The analysis of Fig.~\ref{fig:36} allows such an
interpretation: Towards $PA$ = 120~$^\circ$ {the
  measured visibilities appear to be slightly lower than in orthogonal
  direction, necessitating a } larger extension of the
brightness distribution in this direction.
Such a lateral contraction could be evoked by the strong stellar wind
of a nearby star, but more probably this deviation from circular
symmetry can be attributed to a wavelength dependent size (cf. Sect.~\ref{sec:32}). 

\subsection{Radiative transfer models}
\label{sec:47}

{The Gaussian model analysis of the data demonstrated that the interferometer resolves the inner dust around IRS~3.
Thus we apply a physically self-consistent radiative transfer (RT)
model  here to further investigate the physical and chemical properties of the inner dust around IRS~3 addressed before by the smaller of the two Gaussians.}

Because of the circular symmetry of the source, we use the one-dimensional
code {\sc Dusty} \citep{1999dusty}. 
Since IRS~3 appears to be very luminous, isolated, and surrounded
by a lot of dust, it is the most reasonable to assume that IRS~3 is a post
main-sequence star with strong stellar winds and massive dust formation. 
We followed a heuristic approach and calculated four distinct scenarios spanning the space of possible
stellar parameters: {\em hot} and {\em cold}, realized by stellar
effective temperatures of $T_{\rm *,\,hot}=2.5\cdot10^4~$K and $T_{\rm
  *,\,cold}=3\cdot10^3~$K; {\em C-rich} with C/O abundance ratios beyond 1,
realized by a domination of amorphous carbon grains in the circumstellar dust and {\em O-rich} with a dust composition dominated by warm amorphous silicates.

We applied radial density profiles dominated by radiation pressure of
the star \citep{1995ApJ...445..415I}.
While the chemical composition and temperatures of the dust strongly influences the
infrared spectrum, the stellar effective {temperature $T_{\rm *}$,
  and luminosity $L_*$ scale the physical size of the
system in dependence on $T_{\rm dust}$, the temperature of the
sublimation zone, where dust distribution starts}.
For the grain size distribution we used a power-law distribution as
published by \citet{1977ApJ...217..425M} and an upper limit on the
grain size of $a=0.25~\mu$m, which was successfully applied in
similar experiments.

{To demonstrate the different constraints imposed by the data we present our results in two parts. First the measured SED can be related to the optical depth $\tau$ of the dust and} to the temperature $T_{\rm dust}$ at
the inner boundary of the dust shell. 
{In the next step the remaining ambiguities regarding the stellar temperature and dust composition can be scrutinized further by comparing calculated visibilities of the RT models to the MIDI data.}
 
\subsubsection{Modeling the spectral energy distribution}
\label{sec:84}

\begin{figure}
\centering
\includegraphics[width=\columnwidth]{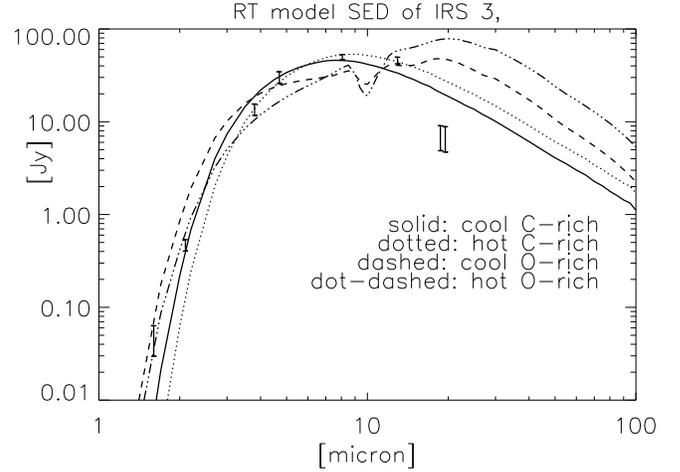}
\caption{{Data equal to the one in Fig.~\ref{fig:45}, but further corrected to match the {\em intrinsic} emission of IRS~3 and its surrounding circumstellar dust shell. The corrections are discussed in the text, and lead to the following intrinsic fluxes:  0.05, 0.5, 13.5, 30, 50, 46, 7.0~Jy at 1.6, 2.1, 3.8, 4.7, 8, 13, 19~$\mu$m. The best RT model SED were overplotted, the respective optical depths $\tau_{13}$ are 0.4 for both C-rich models and 1.6, and 1.3 for the cold and hot star oxygen-dominated models, respectively.
}}
\label{fig:90}
\end{figure}

{This part is aggravated by the fact that IRS~3 is observed both through a large amount of interstellar extinction towards the GC and through the local (partly diffuse) dust of non-negligible optical depth at MIR wavelengths as shown in earlier sections. 
Some uncertainty derives from the fact that the shape of the interstellar extinction only at wavelengths shorter than 8~$\mu$m is well observed and shown to be relatively constant throughout the Galaxy \citep{2005ApJ...619..931I}.
In addition we do not know the exact dust composition and resulting shape of the interstellar extinction at MIR wavelengths within the central parsec and around IRS~3 in particular.
Thus the dereddened SED as the basis of this section is expected to create uncertainties on the order of magnitude of tens of percent in flux.

This naturally affects the accuracy of the estimation of $T_{\rm dust}$ and $\tau$, which both determine the shape of the intrinsic SED.}
 Similarly we cannot meaningfully investigate the gradual importance of secondary ingoing parameters like the density profile,
chemistry (e.g. existence and amount of SiC, crystalline olivines), etc.
These parameters cannot be fine-tuned
unambiguously by comparison with the dereddened stellar spectrum, so fixed standard values are used. 

Since the {\em relative} depth of the interstellar silicate
absorption is unclear, we excluded the data around 9.8~$\mu$m for the SED fits and used only data at the {edges} of the $N$-band. 
{In addition we corrected the SED for using only 40, 30, and 20~\% at 8, 13, and 19~$\mu$m of the single telescope flux making allowance for the found interferometric flux ratios, thereby eliminating the emissive contributions from the outer cooler dust. 
Furthermore, we increased the optical depth at $\lambda \ge 8~\mu$m beyond the $A_{\rm V}=25$ seen at NIR wavelengths to account for the extinction caused by the outer flux seen by the interferometer (Sect.~\ref{sec:44}).
This is realized by upscaling the assumed shape of the interstellar extinction towards the GC \citep{2001A&A...366..106M} to match at $\lambda \ge 8~\mu$m an additional $\tau_{13} \approx$ 0.5 (Table~\ref{tab:44}).
This opacity would already explain at least half the $\tau_{9.8}$ excess (Fig.~\ref{fig:33}) if the same interstellar GC extinction shape would again be used for the outer dust addressed by the larger Gaussian in our simple two-component model.
It is conceivable that, by confining this correction to $\lambda \ge 8~\mu$m, we slightly underestimate the dereddened fluxes at shorter wavelengths, which should also be affected by the outer dust.

The remainig excess extinction at 9.8~$\mu$m could stem from either fractionally enhanced silicate dust in the central parsec in general or in the outer and (partially diffuse) dust around IRS~3 only. 
In the next section we aim at investigating whether this remaining  $\tau_{9.8}$ excess towards IRS~3 could originate from the inner dust at 20~mas scales and is thus produced in an oxygen-rich dust shell. 
Finding the latter would suggest IRS~3 to contribute to its possibly silicate-enriched environs.}

{We decided to fix $T_{\rm dust}$ for all models at a typical value for subliming dust  of 1200~K. 
Tests show that by varying this $T_{\rm dust}$ by up to 300 K, we still reproduce the earlier dust temperature findings and generate model SEDs with deviations from the data that can easily be explained by slightly varying the optical depth in the outer dust of the stellar surrounding. 
However, sublimation temperatures well below 900~K can be excluded.
Furthermore, the optical depth needed to create model SED that comply with the data is not negligible.
This in turn enables us to scrutinize the dust composition in the next section since the different optical properties of carbon and oxygen-rich dust show stronger impact in the case of higher optical depths.

The probed optical depth range covered two orders of magnitude ($\tau_{13}=0.1..10$).
The best-fit SED models under the given assumptions are shown in Fig.~\ref{fig:90}. 
The main results can be summarized as follows:
\begin{itemize}
\item the O-rich SEDs show much poorer coincidence with the dereddened data at longer wavelengths;
\item the MIR optical depth of the O-rich models is with $\tau_{O,\,13} \sim 1.5$ even outside the silicate feature that significantly larger than for C-rich dust ($\tau_{C,\,13}\sim 0.4$); this is the only finding that would comply with a scenario in which the $\tau_{9.8}$ excess originates in the innermost dust;
\item the photometry data alone can be fitted convincingly by
  {\em single}-shell dust models, and for each dust chemistry both hot {\em and} cold star models show fits of similar goodness
\end{itemize}
} 

\subsubsection{Modeling the visibilities}
\label{sec:85}

\begin{figure}
\centering
\includegraphics[width=\columnwidth]{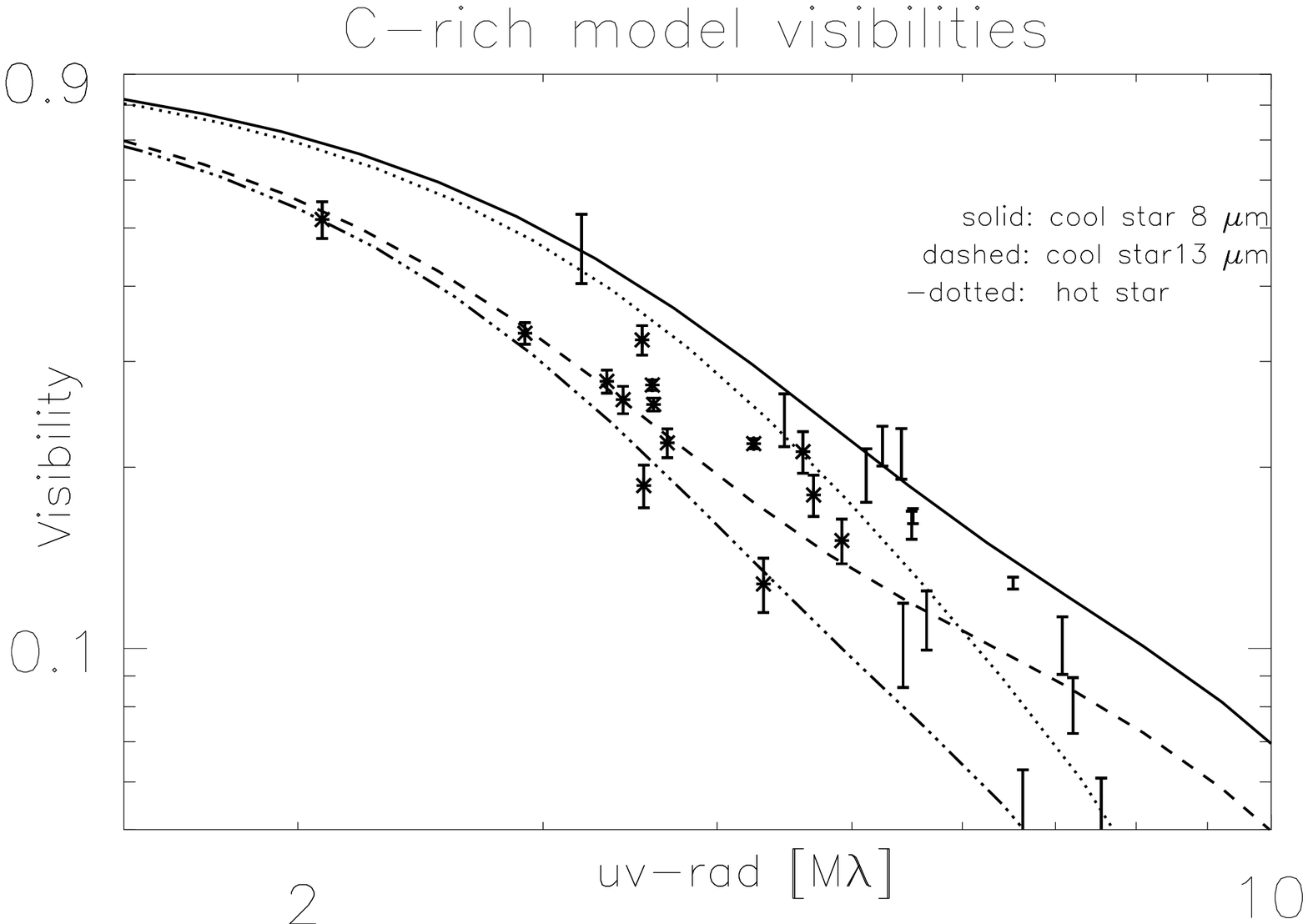}
\includegraphics[width=\columnwidth]{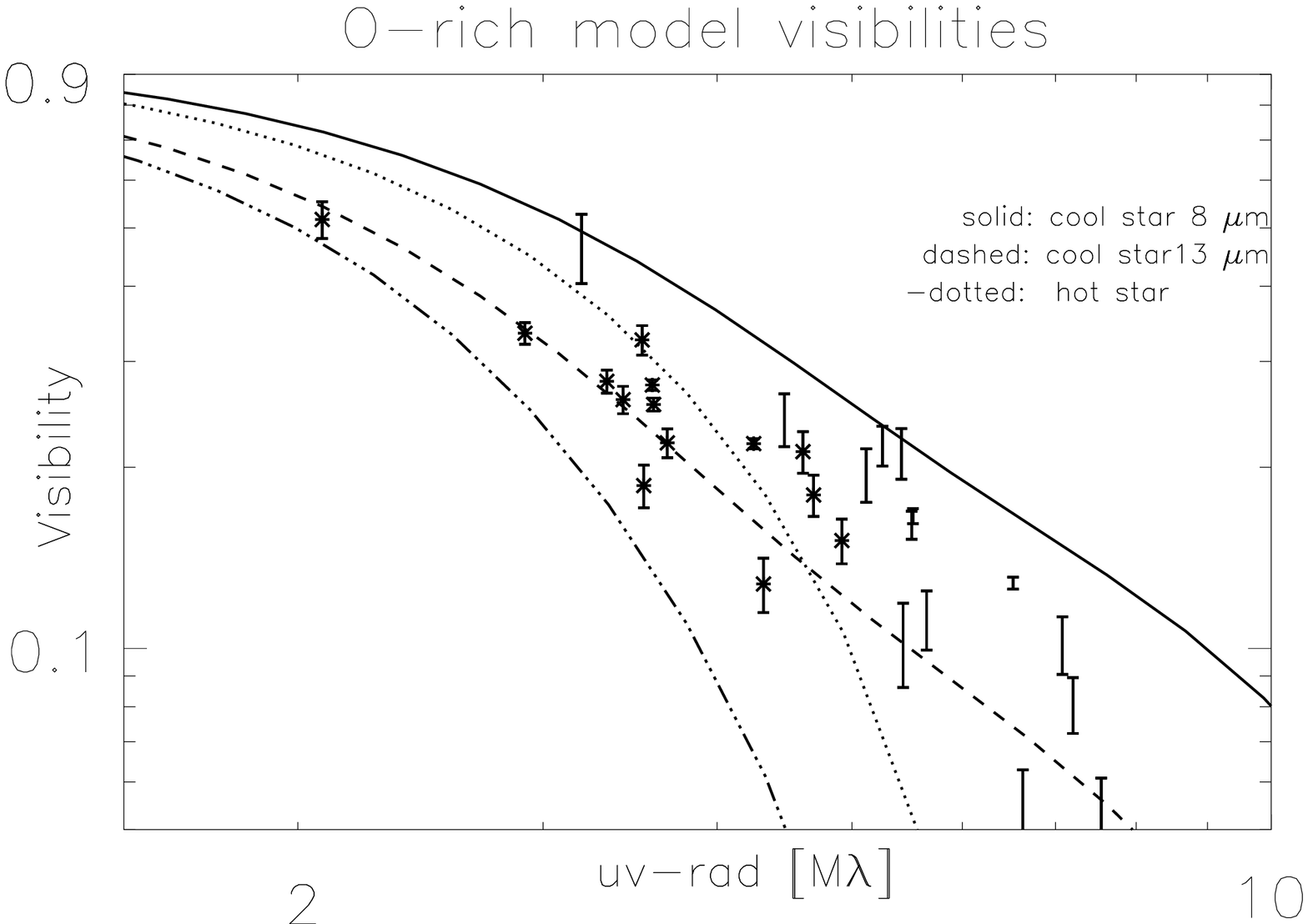}
\caption{{Measured visibilities on the edges of the $N$-band outside the silicate feature. The data at 13~$\mu$m are highlighted by stars. The respective visibilities of RT models plus outer Gaussian (see text) are overplotted as indicated. Quantitative details of the RT models are given in Table~\ref{tab:50}}}
\label{fig:91}
\end{figure}

\begin{figure}
\centering
\includegraphics[width=\columnwidth]{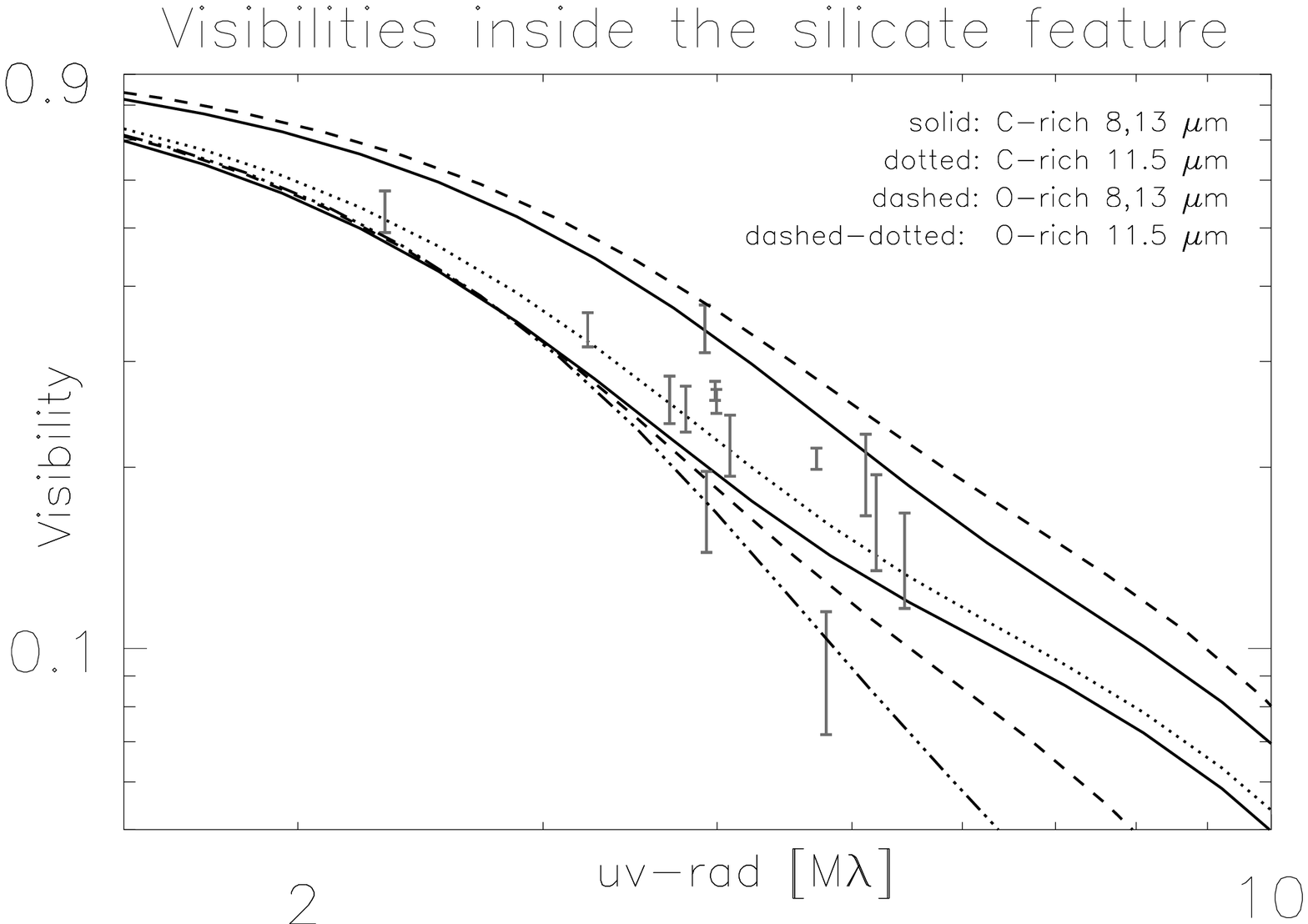}
\caption{{Measured visibilities at 11.5~$\mu$m in the wing of the silicate feature. The best cool star RT models plus Gaussian are overplotted, and the dotted lines represent the models at 11.5~$\mu$m. While the C-rich model at 11.5~$\mu$m lies between the model visibilities at 8 and 13~$\mu$m, as does the data, the O-rich visibilities at 11.5~$\mu$m are lower than ones on the edge of the $N$-band at similar baseline lengths. }}
\label{fig:92}
\end{figure}

\begin{figure}
\centering
\includegraphics[width=0.8\columnwidth]{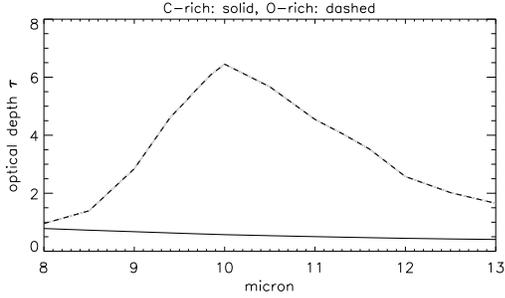}
\caption{Comparison
  of the $N$-band optical depth of cool star models showing the significant change in optical depth along the silicate features of O-rich dust.}
\label{fig:54}
\end{figure}

\begin{figure}
\centering
\includegraphics[width=\columnwidth]{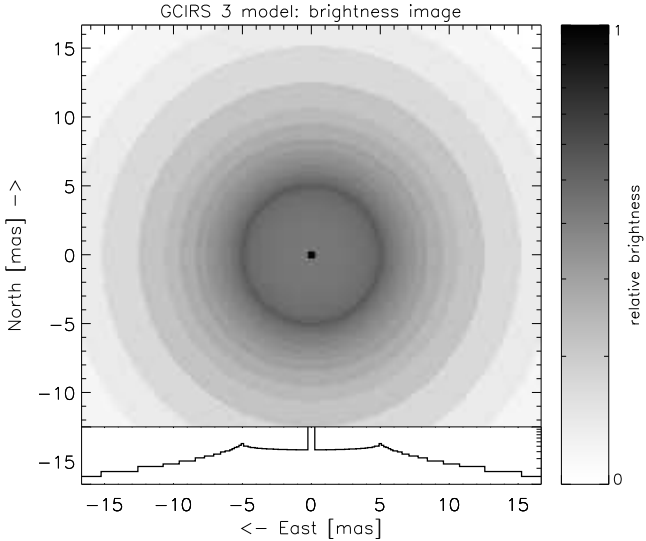}

\caption{Mean $N$-band surface brightness distribution of our cool carbon star
model of IRS~3. 
The inner rim, the dust sublimation zone, is visible as an annulus and resolved by fitting the RT model to our MIDI data. 
At the bottom the radial profile of the respective image is shown, and the image and radial plot are equally scaled logarithmically.}
\label{fig:61}
\end{figure}

\begin{table}
\begin{minipage}[t]{\columnwidth}
\caption{Best parameters of the RT models, shown in Fig.~\ref{fig:90}~-~\ref{fig:61}. The dust shell was confined to end at a radius of
  $10^4\,R_{\rm i}$.
}
\label{tab:50}
\centering         
\renewcommand{\footnoterule}{}  
\begin{tabular}{cccccc} 
\hline\hline       
$T_*$ \footnote{$T_*$ denotes the
  stellar effective temperature} & $T_{\rm dust}$ \footnote{$T_{\rm dust}$ is the dust temperature
  at the sublimation zone, visible as ring of radius $R_{\rm i}$ in Fig.~\ref{fig:61}.}& Comp. & $\tau_{8.0, 9.8, 13.~\mu {\rm m}}$ & $R_{\rm i}$ \footnote{$R_{\rm i/*}$ is the size ratio of the stellar and dust sublimation radius $R_{\rm i}$.} & $L_{*,\,{\rm bol}}$ \footnote{$L_{*,\,{\rm bol}}$ labels the bolometric luminosity of the embedded star.}\\
 $[$kK$]$ & $[$kK$]$ & & & $[$AU/mas/$R_{i/*}]$ &$[ L_\odot ]$ \\
\hline
3 & 1.2 & amC \footnote{\label{foot:10}100\% amorphous Carbon grains
 \citep{1988ioch.rept...22H}} & 0.8,0.6,0.4 & 35/5/10  & 5$\cdot 10^4$\\
3 & 1.2 & Sil \footnote{\label{foot:11}100\% warm amorphous silicate grains
 \citep{1992A&A...261..567O}} & 1,6,1.7     & 40/5/10 & 5$\cdot 10^4$\\
25 & 1.2 & amC \textsuperscript{\ref{foot:10}} & 0.8,0.6,0.4 & 60/8/1,000  & 5$\cdot 10^4$\\
25 & 1.2 & Sil \textsuperscript{\ref{foot:11}} & 0.8,4.9,1.3 & 115/15/2,000& 5$\cdot 10^4$\\
\hline             
\end{tabular}
\end{minipage}
\end{table}

{Similar to the flux correction in Sect.~\ref{sec:84} we have to alter the theoretical visibilities of our RT models here to account for the outer flux probed predominantly by the short baselines prior to comparing the model to the data.
To conform this correction to the findings of the previous sections, we add to the respective RT model a Gaussian component of twice the RT model flux and of FWHM of 35, and 45~mas at 8, and 13~$\mu$m, respectively. 
Again these parameters reflect average properties of the dust emission at lower spatial frequencies and are variable within the ranges implied by the findings of Sect.~\ref{sec:5}.}

With the given data it remains unclear how much of the outer flux is indeed present in a second shell at the 40~mas spatial scale and how close this dust emission resembles a Gaussian brightness distribution. 
The apparent existence of such a second component might be an artefact of our relatively simple RT models. 
Similar apparently multi-component dust shells around the well-studied cool carbon star \object{IRC+10216}  could be resolved into one complex dust shell by more detailed modeling \citep[e.g.][]{1996MNRAS.279.1019I,
  2001A&A...368..497M}. 
But it was demonstrated that our approach reflects the data and the spatial information therein.
Moreover, the spectral shape of the higher spatial frequencies of the innermost dust discussed here is not significantly influenced by this outer dust.
A variable mass loss rate of the central star might be responsible for our not observing a simple dust morphology that could be explained by a single simple dust shell alone.

{Further it is important to understand that any wavelength-dependent interstellar extinction only decreases the flux but not the flux-normalized visibility, implying that the following results are less affected by the uncertainty in the interstellar MIR extinction along the line of sight to IRS~3.} 

The investigation of the four best model scenarios  fitted to the dereddened SED lead to two important results. 
First the aforementioned ambiguity of $T_*$ is probed by our interferometric data.
The $R_i$ of hot stars with the assumed $T_{\rm dust}^{fit}$ tend to be too large implying visibilities too low to model the measurements (Fig.~\ref{fig:91}~\&~Table~\ref{tab:50}). 
{Increased $T_{\rm dust}$ could alleviate this size problem of the hot star models, and the described uncertainties in the derivation of the intrinsic fluxes and visibilities of IRS~3, as well as the scatter in the data, prohibit a definite answer for the stellar temperature.}
But the comparison of the fitted dust sublimation radii (Table~\ref{tab:50}) with the simple Gaussian analysis in Sect.~\ref{sec:5} also favors the cool star models since such Gaussian fits tend to show FWHM twice as large as the sublimation diameter of respective RT models \citep{1996MNRAS.279.1011I}.
{And dust sublimation radii of 30-40~AU with stellar luminosities of $5\cdot 10^4~L_\odot$ found for the cool star models are common values for dust-forming AGB stars.}
Thus for the first time we can present an experimental indication of the exclusion of WR-like hot star scenarii for IRS~3. 

The second result of this section rules out silicates as a dominating constituent of the circumstellar dust shell.
{The calculated visibilities of the competing
  O-rich and C-rich dust compositions are rather similar at 8 and
13~$\mu$m, far away from the
central silicate absorption, although already here the smaller size difference of the C-rich models between both wavelengths are closer to the wavelength dependence of the data (Fig.~\ref{fig:91}~\&~\ref{fig:92}).
However, towards the center of the silicate feature, the apparent source size of O-rich dust shells increases significantly due to the increase in optical depth (Fig.~\ref{fig:54}). 
Figure~\ref{fig:92} shows that this effect is still visible at 11.5~$\mu$m where we have precise visibility measurements. 
A significant size increase (and visibility drop) at the wings of the silicate feature as predicated by all O-rich dust-shell models that are close to our data has not been observed.
Actually the visibility data do not reveal any clear spectral feature but instead shows a smooth wavelength dependence within the calibration uncertainties (Fig.~\ref{fig:31}).

{Summarizing  the RT analysis of our data, the most probable scenario for the nature of IRS~3 is a cool dust-forming carbon star.}
In Fig.~\ref{fig:61} we have plotted a mean $N$-band brightness distribution
of our best-fit, C-rich cool star model.
The inner rim of dust sublimation, resolved by our interferometric experiment,  is clearly visible at a radius of 5~mas in the brightness map.
{The estimated stellar luminosity (Table~\ref{tab:50}) points
to a cool carbon star on the AGB. Recent studies of carbon stars find similar stellar and circumstellar dust properties \citep{1995A&A...293..463G}.}
The reasonable assumption of a stellar wind driven dust
density distribution, valid for windy post-main sequence stars,
enables us to  calculate the mass loss $\dot M$, terminal
outflow velocity $v_\infty$, and an upper limit on the stellar mass $M_*$
\citep{1996MNRAS.279.1019I}.
We obtain the
following results for the carbon-rich cool star set of parameters
(cf. Table~\ref{tab:50}), which describes the observations best: 
\begin{eqnarray}
\dot M\,=\,6\cdot 10^{-5}\,M_\odot\,{\rm yr}^{-1};\quad
  v_\infty\,=\,30\,{\rm km~s}^{-1}\, .
\end{eqnarray}
The calculations are based on a gas-to-dust mass ratio of 200 and a dust grain bulk density of
  3~g~cm$^{-3}$.
The inherent uncertainties are discussed by \citet{1999dusty}.

The respective optical depth of the carbon-rich dust shell in the NIR
is $\tau_{\rm 2\,\mu m} =6$, confirming that
the circumstellar dust is optically thick in the NIR, as already suggested by
the analysis of the spatially unresolved SED
(Sect.~\ref{sec:46}).  
Thus photospheric CO bandheads, typical of red giant NIR spectra,
cannot be observed, which explains the impossibility of spectral classification
of the central object as described by many authors \citep[most
  recently][]{2006ApJ...641..891T}.

\section{Conclusions}
\label{sec:6}

{We have presented new high-resolution 10~$\mu$m data of the enigmatic
IRS~3 object in the immediate vicinity of SgrA*. The analysis and
interpretation of the data contribute to the understanding of both the nature
of the embedded star and the location of the unusually strong interstellar
silicate absorption towards IRS~3.  The new high-resolution VISIR
imaging data clearly resolve IRS~3 into both a compact and a diffuse
emission component for the first time. 
Most probably the deep 10~$\mu$m silicate absorption feature towards IRS~3 takes place partially in
the local interstellar dust, which is responsible for the substantial diffuse MIR
emission. 

The interferometric data bear convincing evidence that IRS~3 is a cool star within a carbon-rich dust-shell without significant circumstellar silicate absorption.
The clear imprint in the
circularly symmetric interferometric data of the existence of more extended circumstellar dust was shown.} A one-dimensional radiative transfer model can successfully explain the complete,
available near and mid infrared data, spatially unresolved by single telescope
measurements, if {outer dust emission on 40~mas scales and beyond and non-negligible additional optical depth is accounted for}.
We estimate
a radius of 35~AU and a temperature of $\sim\,1200~$K for the inner dust rim around IRS~3.

With a bolometric luminosity of 5~$\cdot 10^4\,L_\odot$ and a
stellar temperature of about 3000~K IRS~3 appears to be a cool
luminous carbon star, {most probably in the helium-core burning phase. 
Such mass-losing stars are important for  enriching their environment. 
{Only because of the high angular resolution of the interferometer could we demonstrate that a large fraction of the dust is produced by IRS~3 itself, whereas most other MIR bright sources in the central parsec have been shown to heat the surrounding interstellar dust that has already formed \citep{2005ApJ...624..742T,2006JPhCS..54..273P}. Thus IRS~3 appears to be the prime candidate for observing ongoing dust formation in the immediate vicinity of a supermassive black hole.}}

In addition to the interstellar silicate absorption, we found
indications of SiC absorption. {For the first time the new
  generation of MIR spectrometers at 8m class telescopes offer the
  possibility of studying in detail the spectral properties of the
  interstellar absorption towards GC stars at angular
  resolution high enough to account for source confusion and substantial diffuse
  emission in this outstandingly dense and dusty region. Since a C-rich circumstellar
  dust shell is free of any silicate feature and IRS~3 is embedded in
  an exceptionally large amount of ISM, the resulting deep absorption
  of the smooth background continuum of IRS~3 appears well-suited to studying the shape of the
  interstellar absorption towards the GC and the extraordinary
  strength of the silicate absorption feature in more detail. The absorption profile
  depends on the chemical composition and is needed to accurately
  correct for the ISM extinction, yielding the intrinsic spectra of
  all GC sources.}

On the basis of our results we propose to grant IRS~3 a central role in such new studies of the
interstellar silicate absorption towards the GC. Furthermore,
a longterm MIR variability study is proposed to investigate periodical
dust formation around IRS~3, suggested by this complex dust morphology. 

\begin{acknowledgements}
We are grateful for many fruitful discussions with A. Zijlstra and several members of
the MIDI consortium, in particular W.~Jaffe, R.~K\"ohler, C.~Leinert,
and T.~Ratzka. 
Detailed comments from the referee helped to make the paper clearer.
The outstanding support of the ESO Paranal VLTI team, guaranteeing
efficient technically advanced observations, is acknowledged.
This research has made use of the SIMBAD database,
operated at the CDS, Strasbourg, France. Part of this work was supported
by the German {\em Deutsche Forschungsgemeinschaft} (DFG via SFB 494).
JUP was funded by an ESO
studentship and appreciates the hospitality of the UCLA GC group. 
\end{acknowledgements}

\bibliography{6733}
\bibliographystyle{aa}

\end{document}